\documentclass[aps,prb,twocolumn,amsmath,amssymb,superscriptaddress,floatfix]{revtex4-1}

\usepackage{amsmath}
\usepackage{amssymb}
\usepackage[toc,page]{appendix}
\usepackage{mathtools}
\usepackage{amsfonts}
\usepackage{graphicx}
\usepackage{comment}
\usepackage{bm}
\usepackage[T1]{fontenc}
\usepackage[colorlinks=true,linkcolor=blue,citecolor=blue,urlcolor=blue]{hyperref}

\DeclarePairedDelimiter\floor{\lfloor}{\rfloor}

\newcommand{\dd}{\mathrm{d}}
\newcommand{\wt}{\widetilde}
\newcommand{\ee}{\mathrm{e}}
\newcommand{\ii}{\mathrm{i}}
\newtheorem{prop}{Theorem}
\begin{document}

\title{The fine structure of heating in a  quasiperiodically  driven  critical quantum  system }

\author{Bastien Lapierre}
\affiliation{Department of Physics, University of Z\"urich, Winterthurerstrasse 190, 8057 Z\"urich, Switzerland}

\author{Kenny~Choo} 
\affiliation{Department of Physics, University of Z\"urich, Winterthurerstrasse 190, 8057 Z\"urich, Switzerland}

\author{Apoorv Tiwari}
\affiliation{Department of Physics, University of Z\"urich, Winterthurerstrasse 190, 8057 Z\"urich, Switzerland}
\affiliation{Condensed Matter Theory Group, Paul Scherrer Institute, CH-5232 Villigen PSI, Switzerland}

\author{Cl\'ement Tauber}
\affiliation{D\'epartement de Math\'ematiques and UMR 8088, CNRS and CY Cergy Paris Universit\'e
	95000 Cergy-Pontoise, France}
	
\author{Titus Neupert}
\affiliation{Department of Physics, University of Z\"urich, Winterthurerstrasse 190, 8057 Z\"urich, Switzerland}

\author{R.  Chitra}
\affiliation{Institute for Theoretical Physics,	ETH Z\"urich,	Wolfgang-Pauli-Str. 27, 8093 Z\"urich, Switzerland}

\date{\today}

\begin{abstract}

We study the  heating dynamics of a  generic  one dimensional critical system  when  driven quasiperiodically.
Specifically, we consider  a  Fibonacci drive sequence comprising  the  Hamiltonian of 
uniform  conformal  field theory (CFT) describing such critical systems  and its sine-square deformed counterpart. 
The  asymptotic dynamics   is dictated by the  Lyapunov exponent which has a  fractal structure  embedding Cantor lines where the exponent is exactly zero.
Away from these Cantor lines, 
 the system typically heats up  fast  to infinite energy      
 in a non-ergodic manner where the  quasiparticle excitations  congregate at a small number of  select spatial locations  resulting in a build up of energy at these points.
 Periodic
dynamics with no heating for physically relevant timescales is seen in the high frequency regime.
As we traverse the fractal region and approach the Cantor lines,   the  heating  slows enormously and the  quasiparticles completely delocalise at stroboscopic times.
Our setup allows us to tune between  fast and ultra-slow heating regimes in  integrable systems.

\end{abstract}

\maketitle

\section{Introduction}

Symmetries and their associated conservation laws are of tremendous help in solving physical systems with many degrees of freedom. This is particularly true for interacting quantum mechanical systems.
Considering lower symmetry systems may, however, not only bring about complications but also allow for qualitatively new kinds of behavior. For instance, the study of systems with broken translation symmetry has brought to light the phenomena of Anderson and many-body localization, ultimately shaking up certain foundational beliefs of quantum statistical mechanics~\cite{Basko_2006,basko2006problem,Nandkishore_2015, gopalakrishnan2019dynamics}.

A symmetry that has long been untouched when studying many-body quantum systems is that of time-translation, leading to the conservation of energy. This negligence may be due to the assumption that generic driven systems will eventually heat up to infinite temperature -- arguably a completely boring state. In recent years, however, a much more nuanced picture of driven quantum systems has emerged, including several scenarios in which systems do not heat up or enter an exponentially long \emph{preheating} phase with oscillatory dynamics.
Most studied are Floquet systems, in which time-translation symmetry is broken to a discrete subgroup by a periodic drive. They have been shown to avoid heating when integrable~\cite{PhysRevLett.112.150401} or when many-body localized~\cite{Abanin_2019, Abanin_2016}, providing a curious link between broken translation symmetry in space and (partially) in time.
So-called time crystals even allow for the spontaneous breaking of time-translation symmetry~\cite{PhysRevLett.116.250401, PhysRevLett.117.090402, Choi_2017, Zhang_2017}.
A central question concerning the absence of  heating in driven systems is regarding its stability, that is, whether it is robust or requires a large amount of fine tuning and can therefore never be observed in practice.

In this work, we study heating in a system with broken time and space translation symmetry. We uncover (i) a fractal phase diagram with lines of vanishing heating surrounded by regions of very slow heating and (ii) heating phases with a particular structure of hot-spots where the energy density increase nucleates. The latter finding demonstrates that even a heating regime can support non-trivial \emph{emergent} structures as a system is driven towards the infinite-temperature fixed point. Our system breaks translation symmetry in space via a smooth deformation of hopping parameters, rather than short-range correlated disorder, and in time due to a quasiperiodic drive, which has also been in the focus of several other recent works that study (the absence of) heating~\cite{PhysRevB.99.020306,PhysRevLett.120.070602, Else_2020, PhysRevB.99.220303, Zhao_2019}.

Studying non-periodically driven, disordered many-body quantum systems is about the hardest setting that can be imagined. In order to make analytical progress, we compensate the lack of time and space translation symmetry, by allowing ourselves access to the infinitely generated conformal symmetry group otherwise. Concretely, we study a driven conformal field theory (CFT), where the time-evolution operator alternates between a uniform $(1+1)$-dimensional CFT and one of its non-homogenous versions known as {\it{sine-square deformation}} (SSD)~\cite{Katsura_2012,Okunishi_2016,Ishibashi_2015,Maruyama_2011,Ishibashi:2016bey,Toshiya_2011}. This setup has been previously studied with a periodic Floquet drive, where it displays a rich phase diagram with both heating and non-heating phases~\cite{wen2018floquet,lapierre2019emergent,fan2019emergent}. 
Our quasiperiodic drive sequence is generated by a deterministic recursion relation that does not contain any periodic pattern. Such a protocol is inspired from quasi-crystals with a quasi-periodicity in space which have been intensively studied in the past \cite{PhysRevLett.53.2477, bellissard1989}.  More precisely, here we study a protocol that alternates between homogeneous CFT and  SSD according to the celebrated Fibonacci sequence. This results in an exactly solvable quasi-periodically driven interacting model.

Focusing on the evolution of the total energy and the Loschmidt echo, we show that the energy (almost) always increases exponentially at large times while the Loschmidt echo decays exponentially. Both quantities are controlled by the same rate, called the Lyapunov exponent $\mathcal L$. Thus, the system generically and unsurprisingly heats up. However, we find that this happens with a remarkably broad range of heating rates, depending on the parameters of the drive. We observe fast heating areas analogous to the Floquet setup, as well as regions where the heating rate is very slow, with $\mathcal{L}$ close to zero. Moreover, there exists a region of the parameter space where $\mathcal L$ is exactly zero, so that the system escapes heating even at infinite times, but this region has a Cantor set fractal structure of zero measure. 
Forming a measure zero subspace, these regions are not directly accessible. However, they are evidenced by very slow heating neighborhoods in parameter space which remain non-heating for all experimentally and physically relevant time scales.

The paper is organized as follows. In Sec.~\ref{Sec:setup}, we set up the Fibonacci quasiperiodic drive while in Sec.~\ref{Sec:method} we collect some technical details related to CFT computations that are employed in the rest of the paper. In Sec.~\ref{dynamicsheating} we describe the dynamical phase diagram constructed from the Lyapunov exponent and compare and contrast different regions therein based on the time evolution of two observables, namely the total energy and the Loschmidt echo. In Sec.~\ref{Sec:fractal}, we map the unitary evolution of our setup to a classical dynamical map known as the Fibonacci trace map and use it to prove that the region of vanishing Lyapunov exponent (non-heating at infinitely long times) forms a measure zero subset of the parameter space. In Sec.~\ref{Sec_hf}, we provide an analytical treatment for the high-frequency regime. In Sec.~\ref{Sec:qp} we discuss the quasiparticle picture and finally describe related numerics in Sec.~\ref{Sec:numerics}. We provide further details on CFT computations of the Loschmidt echo, the Fibonacci trace map, the high frequency expansion and a ``M\"obius" generalization of our quasiperiodic drive in several appendices.

\section{Setup of the Fibonacci drive}
\label{Sec:setup}
 We consider   a spatial deformation of a generic homogeneous $(1+1)$ dimensional CFT with central charge $c$ and of  spatial extent $L$ defined by the Hamiltonian,
\begin{equation}
\mathcal{H}[f]=\int_0^L \dd xf(x) T_{00}(x),
\label{genericdef}
\end{equation}
 where $T_{00}(x)$ is the energy density of the CFT. These inhomogeneous conformal field theories have been studied in the context of quantum quenches\cite{Dubail_2017, Bastianello_2020, Ruggiero_2019, Allegra_2016, kosior2020nonlinear} and out-of-equilibrium dynamics\cite{moosavi2019inhomogeneous, Gaw_dzki_2018}.
 
In terms of the Virasoro generators $L_n$ and $\overline{L}_n$, in the Euclidean framework
with imaginary time $\tau$,
the uniform CFT Hamiltonian defined as $\mathcal{H}:=\mathcal{H}[1]= \frac{2\pi}{L}\left(L_0+\bar{L}_0\right)$ is obtained by taking $f\equiv 1$, and the so-called Sine-square deformation (SSD) is defined as $\widetilde{\mathcal{H}}:=\mathcal{H}[2\sin^2\left(\frac{\pi x }{L}\right)]=\frac{2\pi}{L}\left( L_0-\frac{1}{2}(L_1+L_{-1})+\bar{L}_0-\frac{1}{2}(\bar{L}_1+\bar{L}_{-1})\right)$. The advantage of such sine-square deformation of the CFT is that these theories have been widely studied in the context of dipolar quantization\cite{Okunishi_2016,Ishibashi:2016bey}. Furthermore because of the $SL(2,\mathbb{C})$ structure of such a deformation, the full time evolution with the SSD Hamiltonian can be obtained analytically\cite{Wen_2018}.

\begin{figure}[t]
	\includegraphics[width=8.6cm]{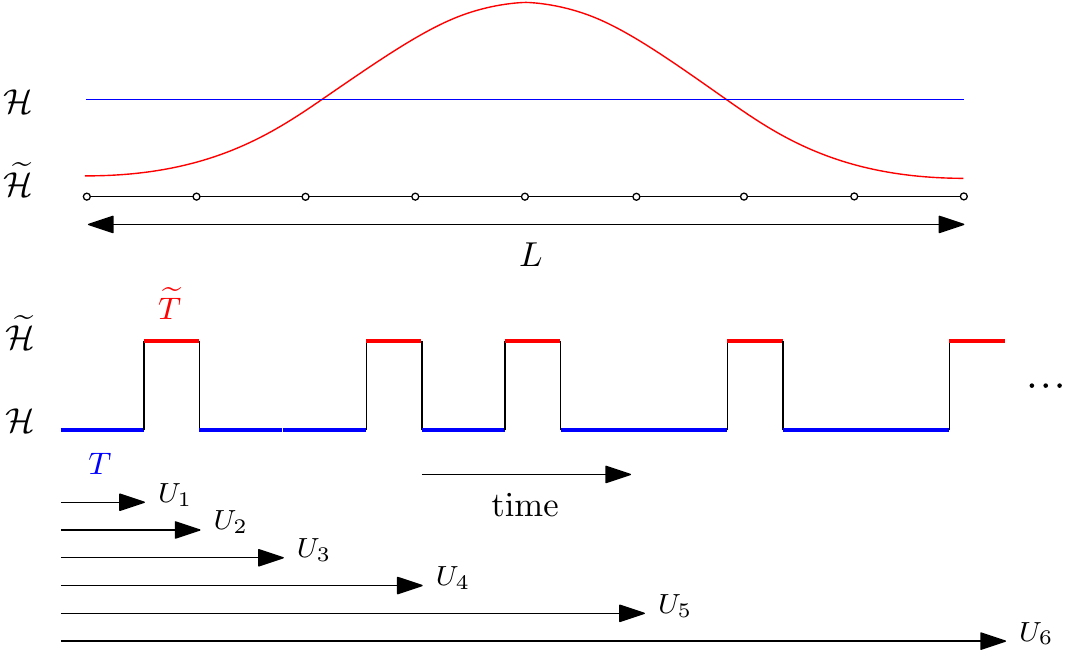}
	\caption{\label{fig:fibodrive} Top: Uniform Hamiltonian $\mathcal{H}$ and  the SSD Hamiltonian $\widetilde{\mathcal{H}}$ 
	Bottom: Fibonacci quasiperiodic drive. The time evolution follows the relation \eqref{timeevofibo}. At every Fibonacci step $n$, the time evolution involves  $F_{n+1}$ unitary operators,  comprising $F_{n-1}$ times $U_0$ and $F_{n}$ times $U_1$. }
\end{figure}

Recently, in Refs.~{\onlinecite{wen2018floquet,lapierre2019emergent,fan2019emergent}},  the Floquet  dynamics of  an interacting critical  field theory  based on a step drive alternating  periodically between  the undeformed $\mathcal{H}$ and the deformed Hamiltonian $\wt{\mathcal{H}}$  was studied.  Based on   a classification of M\"obius transformations which encode  the time evolution of the system over one period via a conformal mapping \cite{wen2018floquet},   a phase diagram comprising both heating and non-heating phases was obtained. The heating was shown to be  related to the emergence of black hole horizons in space-time and inherently non-ergodic\cite{lapierre2019emergent}.  In the  extreme limit  of a  purely random drive,  the system was shown to lead to  always heat up \cite{fan2019emergent}.   A natural question is  what happens  in the intermediate case,  where the driving protocol  is neither periodic nor completely random - i.e., the  case of  \textit{quasiperiodic driving} where  the  drive is determined by a recursion relation, but for which one cannot extract any periodic pattern.   We address this question in the present work.

   A canonical choice is  Fibonacci  driving, where the  relevant  recursion relation  which determines  the driving sequence  is the Fibonacci relation, defined as \cite{PhysRevB.99.020306, PhysRevLett.120.070602}
\begin{equation}
U_{n+2}=U_{n}U_{n+1},
\label{timeevofibo}
\end{equation}
with initial conditions $U_0=\ee^{-\ii \wt{T}\widetilde{\mathcal{H}}}$ and $U_1=\ee^{-\ii T\mathcal{H}}$, where $T$ and $\wt{T}$ are the periods of the stroboscopic steps with respectively $\mathcal{H}$ and $\widetilde{\mathcal{H}}$. Denoting the pulse associated to $\mathcal{H}$ as A and the pulse associated to $\widetilde{\mathcal{H}}$ as $B$, the first few terms in the drive sequence  are $ABAABABAABAAB...$, as illustrated on Fig.~\ref{fig:fibodrive}.  Such a drive is  defined  by  three parameters: $(T,\widetilde{T},L)$. 

 In particular, the number of unitary operators at the step $n$ is given by $F_{n-1}$ for $U_0$ and $F_{n}$ for $U_1$, where $F_n$ is the $n$-th Fibonacci number, giving $F_{n+1}$ unitary operators at the step $n$. Therefore,  the number of operators at the step $n$ grows exponentially with $n$; for large $n$, $F_n$ scales as $\frac{\Phi^n}{\sqrt{5}}$, where $\Phi=\frac{1+\sqrt{5}}{2}$ is the Golden ratio. For practical purposes, we also introduce a "stroboscopic" time which counts the unitary operators one by one, that we denote by $N$. One way to count the evolution operators one by one is to introduce $\nu(N)\in\{0,1\}$, with $\nu(N)=\floor*{(N+1)\Phi}-\floor*{N\Phi}-1$. If $\nu(N)=0$, the unitary operator which appears at step $N$ is $U_0$, and if $\nu(N)=1$ the unitary operator at step $N$ is $U_1$. The main question is to understand whether a non-heating region can still survive in the quasi-periodic drive, or if only heating will exist as in the random case.\\

\section{Methodology}
\label{Sec:method}

The full  time evolution  under the quasi-periodic drive is obtained in a similar way as for the periodic case\cite{wen2018floquet, lapierre2019emergent, fan2019emergent}:  we first note that  in  the Heisenberg picture,  the time evolution of any primary field $\phi(x,\widetilde{T})=\ee^{\ii\widetilde{\mathcal{H}}\widetilde{T}}\phi(x,0)\ee^{-\ii\widetilde{\mathcal{H}}\widetilde{T}}$ amounts to a simple conformal mapping. This can be seen by:  (i)  rotating to imaginary time $t\rightarrow \tau$ and (ii) mapping the space-time manifold to the complex plane with the exponential mapping $z=\ee^{\frac{2\pi (\tau+\ii x)}{L}}$ and  finally,  (iii) using the fact that the time evolution is encoded in a particular conformal transformation of the complex plane, denoted $\tilde{z}_0(z)$.  Following this procedure, the full time evolution of the primary field $\phi$ of conformal weight $h=\bar{h}$ is given by 
\begin{equation}\label{primary}
\phi(x,\tilde{\tau})=\left(\frac{2\pi}{L}\right)^{2h}\left(\frac{\partial \tilde{z}_0}{ \partial z}\right)^{2h}\left(\frac{\partial \bar{\tilde{z}}_0}{ \partial \bar{z}}\right)^{2h}\phi(\tilde{z}_0,\bar{\tilde{z}}_0),
\end{equation}
with $\tilde{z}_0(z)$  given by a  simple M\"obius transformation 
\begin{equation}
    \tilde{z}_0(z)=\frac{(1+\frac{\pi \tilde{\tau}}{L})z-\frac{\pi \tilde{\tau}}{L}}{\frac{\pi \tilde{\tau}}{L}z+(1-\frac{\pi\tilde{\tau}}{L})}.
    \label{tildezeq}
\end{equation}
Similarly, the time evolution, with respect to 
 $\mathcal{H}$  is a  simple dilation in the complex plane, such that in \eqref{primary},   $\tilde{z}_0 \to \tilde{z}_1=e^{\frac{2\pi \tau}{L}}z$.\\
 
Consequently, the {\it Fibonacci time evolution} with the Fibonacci quasi-periodic drive amounts to composing the conformal mappings $\tilde{z}_1(z)$ and $\tilde{z}_0(z)$ following the recursion relation \eqref{timeevofibo}
\begin{equation}
\tilde{z}_{n+2}(z)=\tilde{z}_n\circ\tilde{z}_{n+1}(z).
\end{equation}
Equivalently,  time evolution with the {\it stroboscopic time $N$} can also be obtained via  the recursion relation 
\begin{equation} \label{moebius-str}
\tilde{z}_{N}(z)=\tilde{z}_{\nu(N)}\circ\tilde{z}_{N-1}(z).
\end{equation} 
The group properties of the invertible M\"obius transformations directly imply that $\tilde{z}_N$ is also a M\"obius transformation for any step $N$.  We can then introduce the matrices $M_N$ with unit determinants associated to the conformal transformations $\tilde{z}_N$, such that the stroboscopic time evolution amounts to a sequential multiplication of  $SL(2,\mathbb{C})$ matrices with the recursion relation $M_{N}=M_{\nu(N)}M_{N-1}$, and for the Fibonacci times $n$ such that $N=F_{n+1}$, the relation is $M_{n+2}=M_nM_{n+1}$ where
\begin{align}
M_0=\begin{pmatrix}
1+\frac{\ii \pi \widetilde{T}}{L} & -\frac{\ii\pi \widetilde{T}}{L} \\
\frac{\ii\pi \widetilde{T}}{L} &1-\frac{\ii\pi \widetilde{T}}{L} 
\end{pmatrix}, \
M_1=\begin{pmatrix}
\ee^{\ii\pi T/L} & 0 \\
0 & \ee^{-\ii\pi T/L} 
\end{pmatrix},
\nonumber 
\end{align}
and the general matrix after $N$ steps is denoted by
 \begin{equation}
 \label{matrixM}
 M_N=\begin{pmatrix}
\alpha_N & \beta_N \\
\gamma_N & \delta_N
\end{pmatrix}.
\end{equation}

\begin{figure}[t]
	\includegraphics[width=8.6cm]{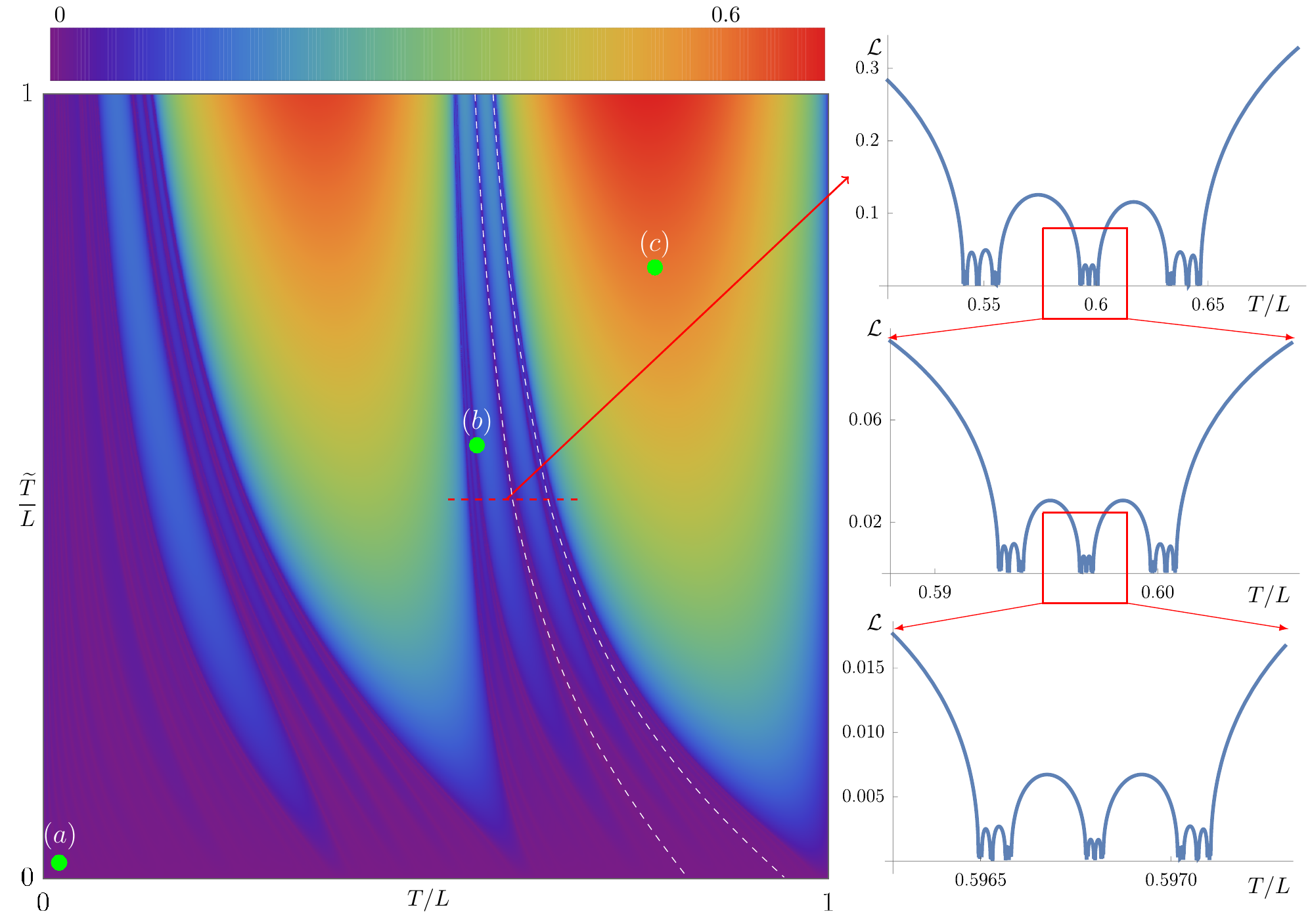}
	\caption{\label{fig:plotlyapunov} Phase diagram obtained from the Lyapunov exponent (Eq.~\eqref{eq:lyapms}) for the Fibonacci drive.    The high frequency regime around point $(a)$ corresponds to non-heating phases whereas the  bright regions, for instance around point $(c)$, correspond to heating phases.  For clarity, only two Cantor lines where the Lyapunov exponent is strictly zero are
	shown in this phase diagram. To elucidate the fractal structure of  the phase diagram,   we  zoom successively across a representative horizontal cut indicated by the red dashed lines.   As we approach the Cantor points embedded in this cut, the system starts to manifest  slow  heating  concomitant with an increasing stroboscopic delocalization of the quasiparticle excitations.   The three marked points marked $(a)$,$(b)$ and $(c)$   represent these three regimes and  will be extensively discussed in the rest of the paper. }
\end{figure}
To address heating  in this quasi-periodic problem, we compute the  time dependent energy density,
 $\mathcal{E}(x,t)=\langle\psi(t)| T_{00}(x)|\psi(t)\rangle$, where $|\psi(t)\rangle$ is the time evolved ground state $|G\rangle$ under the quasi-periodic drive. We choose our initial state $|G\rangle$ to be the ground state of the uniform CFT $\mathcal{H}$ with open boundary conditions. Such a state is in general not an eigenstate of $\widetilde{\mathcal{H}}$, and therefore the time evolution under the drive is non-trivial.  For the sake of clarity, we  now consider  the case $T=\widetilde{T}$.  Using boundary CFT techniques, \cite{fan2019emergent, lapierre2019emergent} the  total energy $E(t)=\int_{0}^L\dd x \mathcal{E}(x,t)$ computed at {\it stroboscopic times}   $t=NT$,    depends solely on the matrix $M_N$ and  takes the following explicit form,  
\begin{equation}
E(t=NT)=\frac{\pi c}{8L}\frac{\alpha_N\delta_N+\beta_N\gamma_N}{\alpha_N\delta_N-\beta_N\gamma_N},
\label{energytoteq}
\end{equation}
where $c$ is the central charge of the CFT.

Another  quantity of interest is the Loschmidt echo $\text{F}(t)$,  which is a measure of revival/coherent evolution in the system.  It is determined by the overlap between the initial ground state $|\psi(0)\rangle=|G\rangle$, and its time  evolved counterpart  $|\psi(t)\rangle$, $\text{F}(t)=|\langle\psi(0)|\psi(t)\rangle|^2$  and can be easily accessed in  the context of boundary-driven CFTs {\cite{PhysRevLett.118.260602}}.  For any  $|G\rangle=\lim_{z,\bar{z}\rightarrow 0}\phi(z,\bar{z})|0\rangle$, where $\phi(z,\bar{z})$ is a primary field of the boundary theory with conformal weights $(\Delta,\bar{\Delta})$, and $|0\rangle$ being the $SL(2,\mathbb{C})$ invariant vacuum, one obtains:
\begin{equation}
\text{F}(t=NT)=\left|\frac{\alpha_N\delta_N-\beta_N\gamma_N}{\delta_N^2}\right|^{2(\Delta+\bar{\Delta})}
\label{loschechoeq}
\end{equation}
The derivation of this formula for the Floquet CFT problem is presented in Appendix~{\ref{applosch}}.
As we will show later,  $E(t)$ and $\text{F}(t)$ are formally related  and this will help a clear characterization of the physics induced by quasi-periodic
driving.
We note that the conformal weights $(\Delta,\bar{\Delta})$ of the primary field generating the ground state with open boundary conditions does not appear in the expression of the energy \eqref{energytoteq}. This is a consequence of the fact that $\langle G| T(z)|G\rangle_{\mathbb{H}}$ evaluated on the upper-half plane $\mathbb{H}$, or equivalently on the unit disk vanishes because of rotational symmetry\cite{Calabrese:2009qy}.

\section{Dynamics of heating}\label{dynamicsheating}
As  in dynamical systems \cite{Strogatz2001-STRNDA}, the growth of  stroboscopic total energy or the decay of the Loschmidt echo for a quasi-periodic drive can be characterized by a Lyapunov exponent $\mathcal{L}$  defined by
\begin{equation}
\mathcal{L}=\lim_{N\rightarrow\infty}\frac{1}{N}\log\text{Tr}(M_N)^2.
\label{eq:lyapms}
\end{equation}
Equivalently,  the  corresponding exponent  for the  Fibonacci time reads $\mathcal{L}=\lim_{n\rightarrow\infty}\frac{1}{F_{n+1}}\log\text{Tr}(M_n)^2$.  As we will show,  if the Lyapunov exponent  $\mathcal{L} > 0 $, then the system will heat, and the heating rate is precisely given by $\mathcal{L}$.  Since the structure of the matrix $M$ is known, the Lyapunov exponent can be numerically computed for all $T, \widetilde{T}$, for a sufficient large number of iterations $N$. 

 For the Fibonacci quasi-periodic drive,  the Lyapunov exponent traces the phase diagram  shown in Fig.~\ref{fig:plotlyapunov}. Different regions emerge, some of them correspond to a strong heating with high Lyapunov exponent, whereas other regions display a fractal structure and rather  small values of the  Lyapunov exponent.  This raises  the following  questions:
  are these two regions heating and if yes,  are they heating the same way?   To  answer these,   we  explicitly compute the stroboscopic evolution of the total energy $E(t)$ and the Loschmidt echo $\text{F}(t)$ using Eqs.~{\eqref{energytoteq}} and {\eqref{loschechoeq}}.

\begin{figure}[htb]
	\includegraphics[width=8.6cm]{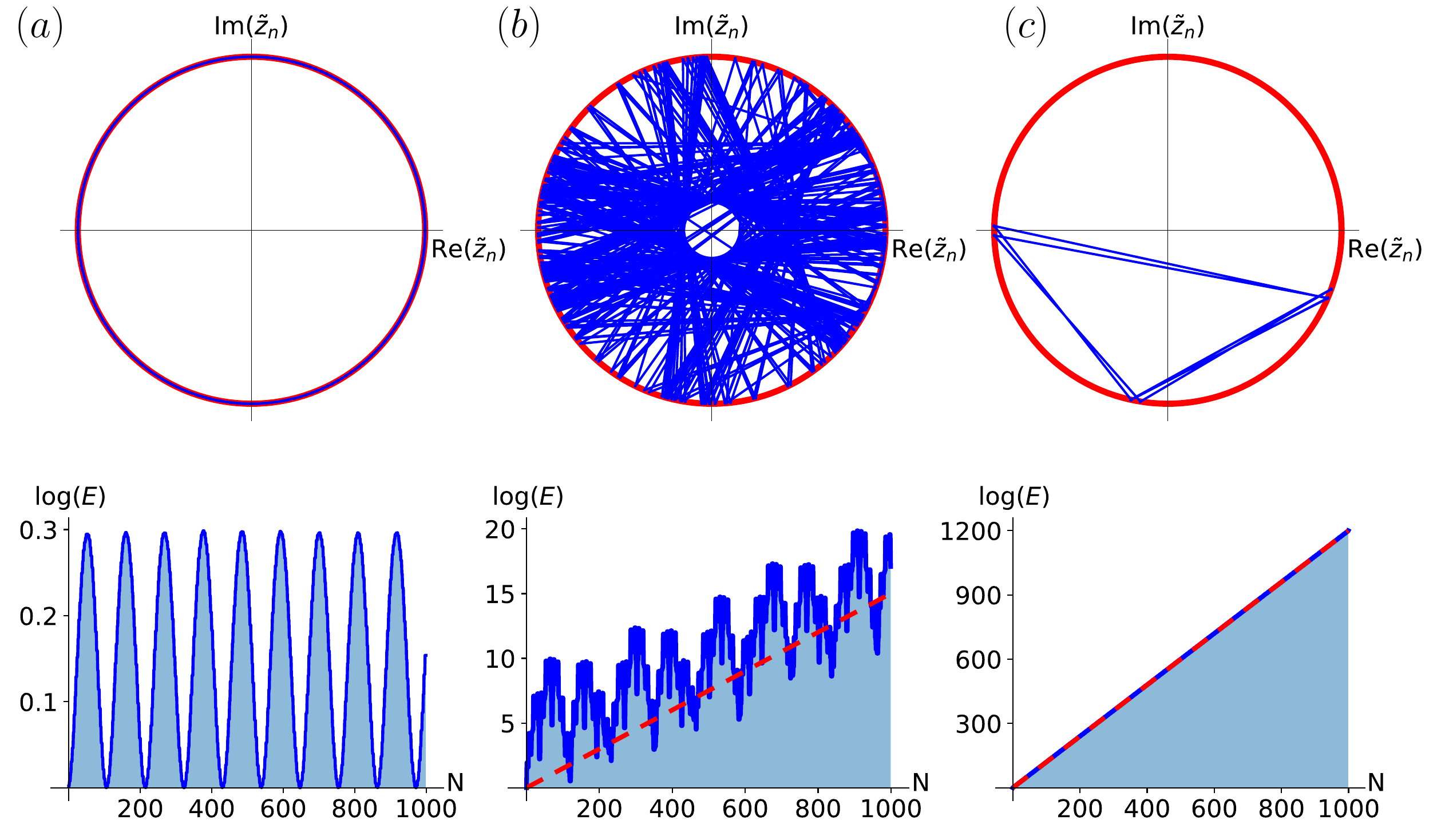}
	\caption{\label{fig:energypluslosch} 
	Top:  Flow of  the conformal mapping $\tilde{z}_N$  when driven by the Fibonacci sequence  with $T=\tilde T$ for upto $N=1000$ and $L=100$. With reference to the phase diagram: $(a)$ $T=1$, in the high-frequency regime the system escapes heating and such a flow is dense on the unit circle. $(b)$ $T=55.101$, in the slow heating regime, the excitation alternate between a large number of recurring regions. $(c)$ $T=78$, as the system strongly heats up, the excitations localize in the system and $\tilde{z}_N$ only has a few recurring points.
	Bottom: 
	Stroboscopic time evolution of the logarithm of the total energy $E(t=NT)$, as a function of the number of iterations $N$.
	$(a)$ in the non-heating  high-frequency regime the energy only oscillates, $(b)$ in the fractal regime  the energy  fluctuates and increases very slowly  and $(c)$ in the heating regime the energy grows exponentially fast with minimal fluctuations. The red dashed lines correspond to the heating rates of the exponentially growing energy at long time, and is given by the Lyapunov exponent $\eqref{eq:lyapms}$.
 }
\end{figure}

First, note that  in  Eq.~\eqref{matrixM}, $\alpha_N=\delta^*_N$ as $M_N \in SL(2,\mathbb{C})$ and because of the form of $M_0$ and $M_1$.  Parametrising  $\alpha_N=R_N\ee^{\ii\phi_N}$, we  obtain  $\text{Tr}^2(M_N)=4R_N^2\cos^2(\phi_N)$ and the  constraint that  $M_N$ has a unit determinant implies, $\alpha_N\delta_N-\beta_N\gamma_N=1$, and $\alpha_N\delta_N+\beta_N\gamma_N=2|\alpha_N|^2-1=2R_N^2-1$.  Using these relations,  the  stroboscopic energy
   $E(t=NT)$ defined in Eq.~\eqref{energytoteq} satisfies,
\begin{equation}
\text{Tr}(M_N)^2=\left[\frac{16L}{\pi c}E(NT)+2\right]\cos^2(\phi_N).
\label{lyapuvsenergy}
\end{equation}
Similarly, the Loschmidt echo  Eq.~\eqref{loschechoeq}  can also be simplified and we obtain  
\begin{equation}
\text{F}(t)=\left|\frac{\alpha_N\delta_N-\beta_N\gamma_N}{\delta_N^2}\right|^{2(\Delta+\bar{\Delta})}=\left(\frac{2}{\frac{8L}{\pi c}E(t)+2}\right)^{2(\Delta+\bar{\Delta})}.
\label{energyvslosch}
\end{equation}

We now establish that the Lyapunov exponent  $\mathcal L$  is indeed the  heating rate in the long time limit. 
From Eqs.~\eqref{eq:lyapms} and \eqref{lyapuvsenergy}, we see that  the Lyapunov exponent 
\begin{equation}
\mathcal{L}=\lim_{N\rightarrow\infty}\frac{1}{N}\log\left(\left[\frac{16L}{\pi c}E(t)+2\right]\cos^2(\phi_N)\right).
\label{relationlyapuenergy}
\end{equation}
Since the oscillatory term $\lim_{N\rightarrow\infty}\frac{1}{N}\log(\cos^2(\phi_N))$ becomes negligible at long times,  we infer that for  an exponential growth of total energy,  the Lyapunov exponent  indeed determines the  heating rate at long times:
\begin{equation}\label{eq:largetime_asymptotics}
\begin{cases}
E(t) \underset{ \overset { t \rightarrow \infty } {} } {\sim} \ee^{\mathcal{L}t/T},\\
\text{F}(t) \underset{ \overset { t \rightarrow \infty } {} } {\sim} \ee^{-2(\Delta+\bar{\Delta})\mathcal{L}t/T}.
\end{cases}
\end{equation}
\noindent

In the case of a periodic drive, the stroboscopic evolution of $E$  and $\text{F}$ show one of three distinct behaviours: (i)  $E$ grows exponentially  and $\text{F}$ decays exponentially with time in the heating phase,  (ii) $E$   and $\text{F}$ oscillate with time in the non-heating phase and  (iii) $E$ grows quadratically  and  $\text{F}$ decays as a power law with time at the transition between the heating and non-heating regimes.   
The results for the quasiperiodic drive are summarised  in Fig.~{\ref{fig:energypluslosch}} where we show three representative scenarios indicated by the dots $(a,b,c)$ in Fig.~\ref{fig:plotlyapunov}.    In the high frequency regime, $T, \widetilde{T}\ll L$ in Fig.~{\ref{fig:energypluslosch}}(a), $\mathcal L$ is very small and   the total energy and the Loschmidt echo oscillate it time akin to the periodic case,  illustrating that  the system avoids heating for a very large number of drive cycles.   In Fig.~{\ref{fig:energypluslosch}}(c),  $\mathcal L$ is large and  we see  standard heating i.e., exponential growth of energy concomitant with an exponential decay of the echo, modulo some oscillations that were not present in the periodic drive case. However,  for the case of $T,\widetilde{T}$ corresponding to Fig.~{\ref{fig:energypluslosch}}(b) where  $\mathcal L$ changes sharply (see  point $(b)$ in Fig.~\ref{fig:plotlyapunov})  new behaviour emerges. We see that  the energy mostly fluctuates and shows very slow growth, while the Loschmidt echo decays slowly and displays strong revivals. 
A fundamental question is then to understand if there exist regions in the phase diagram which can always avoid this exponential growth of energy even at arbitrary long times. We note that at the transition lines $T=kL$ 
, for any $k\in\mathbb{N}$ and any $\widetilde{T}$, $\lim_{N\rightarrow \infty}\tilde{z}_N=1$, and the oscillatory term $\cos(\phi_N)^2\sim \frac{1}{N
^2}$ as $N$ goes to infinity. Therefore this term is not negligible anymore and the energy will grow quadratically even though $\text{Tr}(M_N)^2$ is bounded, thus the Lyapunov exponent is zero. The Loschmidt echo is then decaying quadratically to $0$ as a consequence of Eq.~\eqref{energyvslosch}. Therefore the asymptotic formula \eqref{eq:largetime_asymptotics} is not valid on the transition lines $T=kL$. We note that this quadratic growth of the total energy was already observed in periodic drive at $T=kL$
\cite{fan2019emergent,lapierre2019emergent}, together with a logarithmic growth of entanglement entropy. This dynamics effectively corresponds to a single quantum quench with $\widetilde{\mathcal{H}}$. This can be understood from the quasiparticle picture: if $T=kL$, after the time evolution with $\mathcal{H}$, the quasiparticles will go back to their initial positions. Therefore effectively the system only evolves with $\widetilde{\mathcal{H}}$, implying that all the energy of the system accumulate at the edges of the system, and grow quadratically. 

The  behaviour of the  energy and the echo  are dictated by the 
 stroboscopic evolution of the M\"obius  transformations $\tilde{z}_N$ given by Eq.~\eqref{moebius-str}.  For periodic driving,  the non-heating phase is characterized by a $\tilde{z}_N$ which  oscillates with $N$ and an periodically oscillating energy\cite{lapierre2019emergent}. In the heating phase,  $\tilde{z}_N$ converges to a stable fixed point, $\lim_{n\rightarrow \infty}\tilde{z}_N=\gamma_{1}$, where $\gamma_{1,2}$  are respectively the stable and unstable fixed points of the 1-cycle M\"obius transformation. This in turn  leads to the  creation of two stroboscopic horizons at spatial points $x_c$ and $L-x_c$ determined by the unstable fixed point of the M\"obius transformation, $\gamma_{2}=\ee^{2\pi \ii x_c/L}$ and $\gamma_{2}^{*}=\ee^{2\pi \ii (L-x_c)/L}$ at  which the energy accumulates at  large times.  For quasiperiodic driving  the situation is more subtle.  In the high frequency regime,  $\tilde{z}_N$ traces the unit circle with increasing $N$. The total energy oscillates periodically in this parameter regime [see Fig.~\ref{fig:energypluslosch}(a)]. In regimes where the Lyapunov exponent is large, $\tilde{z}_N$   {\it almost} converges to  a fixed point like scenario but alternates  between a  small set of  points , cf. Fig.~\ref{fig:energypluslosch}(c) resulting in small fluctuations of the energy.   The extreme case  of  this subset comprising only one point, corresponds to the heating  regime of the periodic drive discussed earlier.  As the value  of the Lyapunov exponent decreases and approaches parameter zones where the fractal nature of $\mathcal{L}$ becomes apparent, the flow of $\tilde{z}_N$  becomes more and more dense on the unit circle as seen in Fig.~\ref{fig:energypluslosch}(b), leading to strong fluctuations concomitant with a  very slow growth of the total energy.  It is in these regimes that interesting slow dynamics manifests.

\section{Fractal structure of heating}
\label{Sec:fractal}
We remark that the fractal structure  of $\mathcal{L}$ in  Fig.~\ref{fig:plotlyapunov} as a function of $T,\widetilde{T}$ is very reminiscent of the spectra of a one dimensional Fibonacci quasiperiodic crystal \cite{PhysRevB.35.1020, Kadanoff1276}.  In this section, we will  demonstrate that 
the fractality of $\mathcal{L}$ in our Fibonacci quasiperiodic drive of the CFT  can indeed be related to the spectral properties of  the Fibonacci chain described by the following tight-binding Hamiltonian:
\begin{equation}
H_{mn}=\delta_{m,n+1}+\delta_{m+1,n}+\delta_{m,n}\lambda v(n),
\label{specfibomain}
\end{equation}
where  $v(n)$ is either $0$ or $1$ following the Fibonacci sequence, $v(n)=|\nu(n-1)-1|$.
 
As discussed in Appendix \ref{rouge}, the spectrum $E$ of this Hamiltonian is similar to the Cantor set for any value of $\lambda$.  To see this, we note  that the transfer matrix $T_n$ satisfies the Fibonacci recursion relation, $T_{n+1}=T_{n-1}T_n$.  Since $T_n\in SL(2,\mathbb{C})$, it satisfies the  trace identity \cite{PhysRevB.35.1020}:
\begin{equation}
\text{Tr}(T_{j+1})=\text{Tr}(T_j)\text{Tr}(T_{j-1})-\text{Tr}(T_{j-2}).
\label{tracerelation}
\end{equation}
Introducing $x_j=\frac{1}{2}\text{Tr}(T_{j})$, $y_j=\frac{1}{2}\text{Tr}(T_{j+1})$ and  $z_j=\frac{1}{2}\text{Tr}(T_{j+2})$,   we see that the  trace identity   defines a discrete dynamical map $\mathcal{T}$ called the  Fibonacci trace map,
\begin{equation}
\mathcal{T}:\mathbb{R}^3\rightarrow \mathbb{R}^3,\quad (x_i,y_i,z_i)\mapsto (y_i,z_i,2y_iz_i-x_i).
\label{tracemap}
\end{equation}
The dynamics of a point $ (x_i,y_i,z_i)$  are restricted to a surface defined by the invariant $I(x_j,y_j,z_j)=x_j^2+y_j^2+z_j^2-2x_jy_jz_j-1$.  For the Fibonacci chain,  $I=\frac{\lambda^2}{4}$, and the corresponding set of bounded orbits under the trace map $\mathcal{T}$ for a positive value of the invariant  is  related to the spectrum $E$ of the quasicrystal, which forms a Cantor set of measure zero. 

\begin{figure}[htb]
	\includegraphics[width=8.6cm]{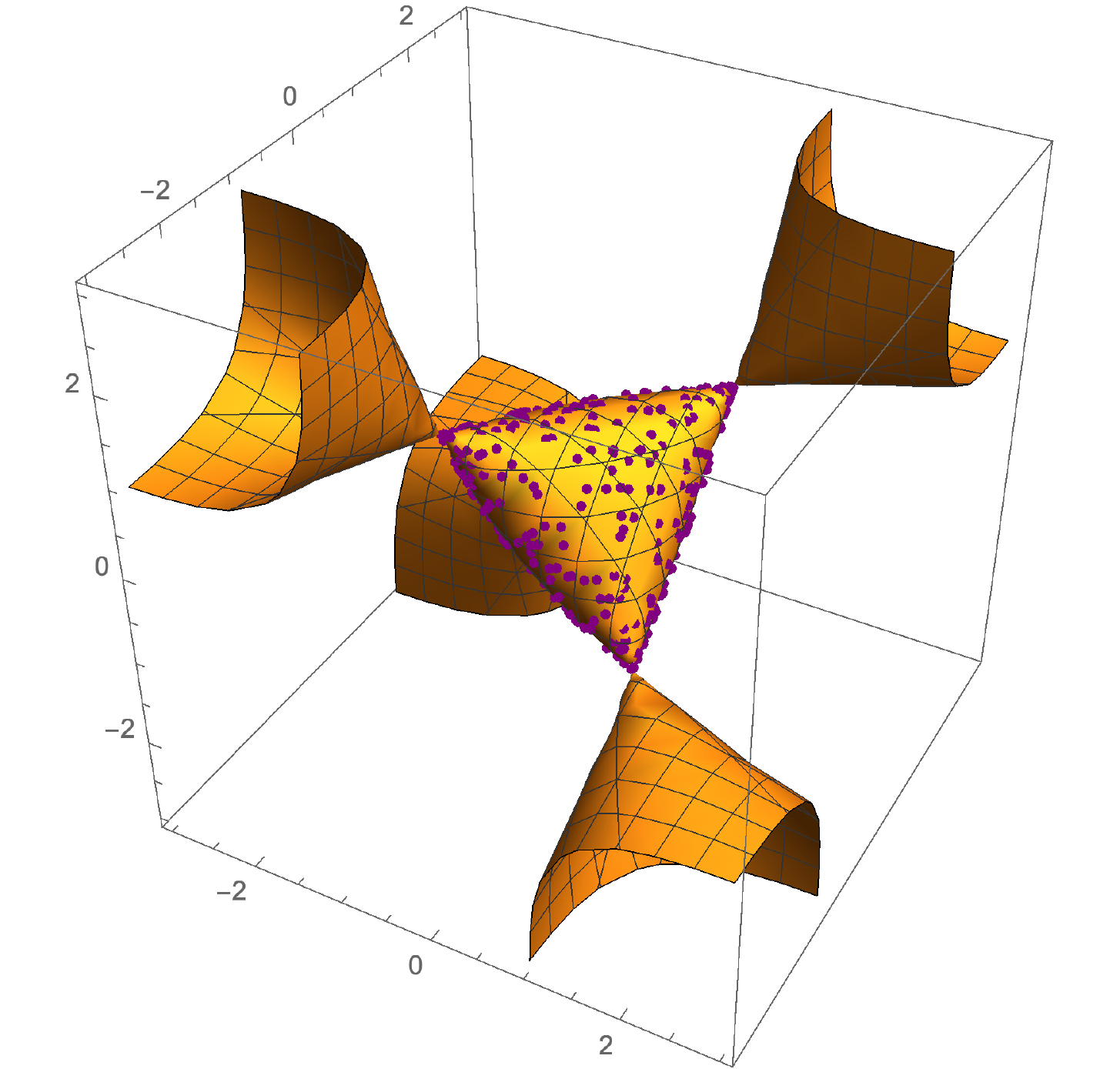}
	\caption{\label{fig:fiboflow} Flow of the Fibonnacci trace map in the high frequency phase, for $L=100$ and $T=\widetilde{T}=1$, for $500$ iterations of the trace map. In this region, the trace map \eqref{tracemap} stays bounded  within the central part of the manifold \eqref{manifold} determined by the invariant \eqref{invarinv}  for a very large number of iterations. In the other regimes, this map will typically be unbounded, apart from a set of parameters of measure 0.
	}
\end{figure}

Based on the fact that  the  time evolution of the quasiperiodic Fibonacci drive is encoded in products of $SL(2,\mathbb{C})$ matrices $M_n$  obeying the Fibonacci sequence $M_{n+1}=M_{n-1}M_n$, we expect $M_n$ to satisfy a  trace relation analogous to \eqref{tracerelation}. This  implies that the trace of the matrix $M_n$ encoding the time evolution after $F_{n+1}$ iterations is completely determined by the orbit of the trace map \eqref{tracemap}. The initial point is simply given by $\left(1,\cos\left(\frac{\pi T}{L}\right),\cos\left(\frac{\pi T}{L}\right)-\frac{\pi \widetilde{T}}{L}\sin\left(\frac{\pi T}{L}\right)\right)$, and the corresponding  invariant is 
\begin{equation}
I=\left[\frac{\pi \widetilde{T}}{L}\right]^2\sin^2\left(\frac{\pi T}{L}\right).
\label{invarinv}
\end{equation}
Clearly this  invariant is always positive, and the associated manifold  explored by the trace map is 
\begin{equation}
\left\{(x,y,z)\in\mathbb{R}^3 \mid I(x,y,z)=\left[\frac{\pi \widetilde{T}}{L}\right]^2\sin^2\left(\frac{\pi T}{L}\right)\right\}.
\label{manifold}
\end{equation}
This manifold is  noncompact and comprises  a central piece connected to four of the eight octants of $\mathbb{R}^3$, as shown  in Fig.~\ref{fig:fiboflow}.  As we iterate the trace map,   orbits typically originate in the central  region of the manifold and  escape   to infinity with a particular escape rate.  As in the Fibonacci chain,  a sufficient condition for an orbit to escape to infinity is that  at some iterative step $j$ \cite{Kadanoff1276}
\begin{equation}
\begin{cases}
|x_j|>1,\\
|x_{j-1}|>1,\\
|x_{j-1}||x_{j}|>|x_{j-2}|.
\end{cases}
\label{conditions}
\end{equation}
A particular case of a bounded orbit is the  trivial fixed point $(1,1,1)$ of the mapping \eqref{tracemap}, which corresponds to the limits $T/L=\widetilde{T}/L=0$ for the Fibonacci drive. This fixed point acts as a saddle point: some points in its vicinity will stay bounded for a very large number of iterations of the trace map, whereas some other points will be strongly repelled,  i.e., their orbit will escape quickly. This  is characteristic of  the high frequency limit $T/L,\widetilde{T}/L\ll1$: if $T>0$, $(1,1,1)$ acts as an attractor and the trace of the $M_n$ stays bounded under $T$ for a large number of iterations, as seen on Fig.~\ref{fig:fiboflow}.   The system  thus avoids heating for  times which are longer than physically relevant timescales. In the case $T<0$, or equivalently taking $\mathcal{H}=-\mathcal{H}[1]$, the orbits diverge away from  $(1,1,1)$ and the system heats up only after a few iterations of $n$.   Consequently,  due to  its  proximity to  the fixed point $(1,1,1)$ of the Fibonacci trace map, we  expect a 
 robust high-frequency expansion for the  Fibonacci quasiperiodically driven problem. A second particular case is the family of transition lines $T/L\in\mathbb{N}$, with arbitrary $\widetilde{T}/L$, for which the initial point is $(1,\pm 1,\pm 1)$, which orbit is bounded. Therefore at those particular lines $\text{Tr}(M_n)$ stays bounded, however as discussed in the previous section the oscillatory term in Eq.~\eqref{relationlyapuenergy} is not negligible anymore and the energy grows quadratically, as for a single quench with the Hamiltonian $\widetilde{H}$.

The  case for arbitrary $T/L$ and $\widetilde{T}/L$ is more subtle. The trace map  discussed above helps us make the  following identifications between the driven case and the one dimensional quasicrystal:
\begin{equation}
\begin{cases}
E=2\cos\left(\frac{\pi T}{L}\right),\\
\lambda=\frac{2\pi \widetilde{T}}{L}\sin\left(\frac{\pi T}{L}\right).
\end{cases}
\label{correspondence}
\end{equation}
Note that in the Fibonacci crystal, the spectrum $E$  of the system  for any positive $\lambda$  is a  Cantor set. This means that for a given $\lambda$, any $E\in\text{spec}(H)\cap[-2,2]$  specifies a  particular value of $T/L$ and $\widetilde{T}/L$ such that  ${\rm Tr}(M_n)$ stays bounded for an \textit{infinite} number of iterations of the trace map, i.e., infinite  stroboscopic  and  Fibonacci  times. The spectrum  $E$ for a given value of the coupling $\lambda$ defines the points in the $T,\widetilde{T}$  phase diagram at which no heating takes place under the quasiperiodic driving at Fibonacci steps $n$, as illustrated on Fig.~\ref{fig:spectrumfibo} in Appendix \ref{rouge}. The non-heating regime forms a Cantor set for a fixed value of $\wt{T}$ and have sub-dimensional line-like locus in the parameter space. We refer to these as the {\it{non-heating lines}}. 
This also explains  the fractal structure of the phase diagram  where the  Lyapunov exponents  approach $0$ \eqref{tracemap}. 
Gaps in $E(\lambda)$ correspond to regions  with high Lyapunov exponents in Fig.~\ref{fig:plotlyapunov},  concomitant with strong heating.
In this regime, the  orbits of the trace map  or equivalently,   $\text{Tr}(M_n)$ diverge super-exponentially once it leaves the bounded central zone.
A correct  numerical evaluation of the Lyapunov exponent \eqref{eq:lyapms}  requires that we consider  $n$  for which $\text{Tr}(M_n)$  has already escaped, i.e., satisfies the conditions~\eqref{conditions}.
To summarize, we see that since non-heating points constitute  a Cantor set of  measure zero, the quasiperiodically driven CFT  will typically  heat up for arbitrary $(T,\widetilde{T},L)$.

\section{High-frequency regime}
\label{Sec_hf}
The   Fibonacci trace map argument  shows that the system will typically heat up  infinitely for any choice of $T, \widetilde{T}$ and $L$, modulo a  Cantor set  which is not physically accessible.   Nonetheless,   there exist regions in the phase diagram   where heating times are so large,  that  for physically relevant times   the system is effectively  in the non-heating regime. An example for such a regime is the high-frequency region of the phase diagram, $T,\widetilde{T}\ll L$.
For quasiperiodic drives, despite the deterministic time evolution  given by Eq.~\eqref{timeevofibo} for all  $(T,\widetilde{T},L)$, it is difficult to  obtain a stroboscopic effective description. This is because   the  approach of the periodic drive  wherein the  $n$-cycle drive  can be recast as a composition of $n$ one-cycle M\"obius transformations  is inapplicable here.  However,   an effective Hamiltonian can indeed be obtained in the high frequency regime, $T=\widetilde{T}\ll L$. 

In the high frequency regime, we  first note that   commutators in the Baker-Campbell-Haussdorf expansion  can be approximated as:
\begin{equation}
e^{-\ii \mathcal{H}T}e^{-\ii \widetilde{\mathcal{H}}T}\approx e^{-\ii T(\mathcal{H}+\widetilde{\mathcal{H}}+\ii \frac{T}{2}[\mathcal{H},\widetilde{\mathcal{H}}])}.
\end{equation}
This simplification enables the  calculation of  an effective Hamiltonian,  $\mathcal{H}_{\text{eff}}^{(N)}$  defined as  $U(N)= \ee^{-\ii NT \mathcal{H}_{\text{eff}}^{(N)}}$, describing the dynamics of the system at stroboscopic time step $N$. 
We reiterate  that  $N$ is the stroboscopic time and not the Fibonacci time.   As  for the periodic drive \cite{lapierre2019emergent},  the $SL(2,\mathbb{R})$ structure of the SSD Hamiltonian, dictates that  $\mathcal{H}_{\text{eff}}^{(N)}$ be some linear combination of $L_0$, $L_{1}$ and $L_{-1}$.
Fixing this linear combination  reduces to  enumerating the number of  times  $\mathcal{H}$, $\widetilde{\mathcal{H}}$, and $[\mathcal{H},\widetilde{\mathcal{H}}]$ appear  in the  time evolution operator  $U(N)$.
Using the approach introduced in Ref.~\onlinecite{PhysRevB.99.020306}, we obtain :
\begin{equation}
\mathcal{H}_{\text{eff}}^{(N)}=\frac{\rho(N)}{N}\mathcal{H}+\frac{\sigma(N)}{N}\widetilde{\mathcal{H}}+\ii \frac{T}{2}\frac{\tau(N)}{N}[\mathcal{H},\widetilde{\mathcal{H}}],
\end{equation}
where
\begin{equation}
\begin{cases}
\sigma(N)=2N-\sum_{m=1}^N(\nu(m)+1),\\
\rho(N)=N-\sigma(N),\\
\tau(N)=\sum_{m=1}^N\left[\nu(m)(m-1)-\floor*{\frac{m\Phi}{\Phi+1}}\right].\\
\end{cases}
\label{somedefini}
\end{equation}
At  every step $N$, the effective Hamiltonian is a $SL(2,\mathbb{R})$ deformation of the uniform CFT. This $f_N(x)$ deformation in the high frequency regime is obtained to be 
\begin{equation} \label{eff-def}
\frac{\rho(N)}{N}+\frac{\sigma(N)}{N}-\frac{\sigma(N)}{N}\cos\left(\frac{2\pi x}{L}\right)+\frac{\pi T}{L}\frac{\tau(N)}{N}\sin\left(\frac{2\pi x}{L}\right).
\end{equation}

As shown in Ref.~\onlinecite{lapierre2019emergent},  if this deformation goes to zero at some locations, heating can occur via  the accumulation of energy at stroboscopic times at these locations.  In this regime, the  quadratic Casimir of $SL(2,\mathbb{R})$ is negative, $c^{(2)}=a^2-b^2-c^2<0$ for a deformation of the form $f(x)=a+b\cos\left(\frac{2\pi x}{L}\right)+c\sin\left(\frac{2\pi x}{L}\right)$.  In the quasiperiodic case,   we find that $f_N(x)$ \eqref{eff-def}  oscillates around the  corresponding  deformation for a  purely periodic drive in the high frequency regime: $f(x)=1-\frac{1}{2}\cos(\frac{2\pi x}{L})$ (see Fig.~\ref{highfreqprof} in Appendix \ref{highfreq}).  Clearly,  since  the deformations remain nonzero,  no heating occurs even at very large times. This is  supplemented by the fact that
 in the high frequency regime  the Casimir invariant $c^{(2)}(N)$  remains positive for large $N$ as long as we stay in the high-frequency regime (see Appendix \ref{highfreq}).
 This agrees with the trace map picture as in the high frequency phase the orbit of the trace map stays bounded for a very large number of iterations as it is in the vicinity of the fixed point $(1,1,1)$.

Heating phases can also be accessed within this high frequency effective hamiltonian formalism.  This can be achieved by considering $\mathcal{H}\mapsto -\mathcal{H}$.   Substituting  $\rho(N)\mapsto -\rho(N)$  in \eqref{eff-def} we obtain the corresponding deformation of the effective Hamiltonian  in this regime.   We again see that,
 with increasing $N$,   $f_N(x)$ oscillates around  that of the corresponding periodic drive $f(x)=-\frac{1}{2}\cos\left(\frac{2\pi x}{L}\right)$ which has horizons at $x_c=L/4$ and $L-x_c$.  
 To summarise,   we have  established  the existence of  parameter regions  where the system avoids heating for any physically relevant timescale.

\section{Quasiparticle picture}
\label{Sec:qp}
In this section, we discuss the physical significance of  the  fractality of the  phase diagram and the associated flows of the M\"obius transformations presented in Sec.~\ref{dynamicsheating}.
As seen earlier,  in the three representative cases for the Lyapunov exponents,  the growth of energy as well as the Loschmidt echo manifest  important differences stemming
from the nature of the flows.
We  will now show that   the structure of these flows are indeed crucial to understanding the nature of the heating.  This is easily done  in the quasiparticle picture, where one can
track the  time evolution of   spatial distribution of the excitations.  First, note that  as we are dealing with a  CFT,  during time evolution with the uniform Hamiltonian $\mathcal{H}$, the excitations which are ballistic trace straight lines  in space-time.  Choosing $T=\widetilde{T}$,   a quasiparticle located at $x_0$ at $t=0$  will reach $\pm x(T)=x_0\pm T$ at $t=T$, where the $\pm$ corresponds to  right and left movers respectively (the velocity has been set to one).  Similarly, for a time evolution with the SSD Hamiltonian $\widetilde{\mathcal{H}}$, the excitations follow null geodesics in a curved space-time determined by the spatial inhomogeneity\cite{Dubail_2017},
\begin{equation}
\pm x_N=\frac{L}{\pi}\text{arccot}\left(\mp \frac{2\pi T}{L}+\cot\left(\frac{\pi x_{N-1}}{L}\right)\right),
\end{equation}
where we denote $x_N:=x(t=NT)$.   The stroboscopic position $x_N$ of the quasiparticles for any initial position $x_0$  can now be obtained  by concatenating the  curved space geodesics and the straight lines  in a sequence fixed by the Fibonacci drive. 

\begin{figure*}
	\includegraphics[width=\textwidth]{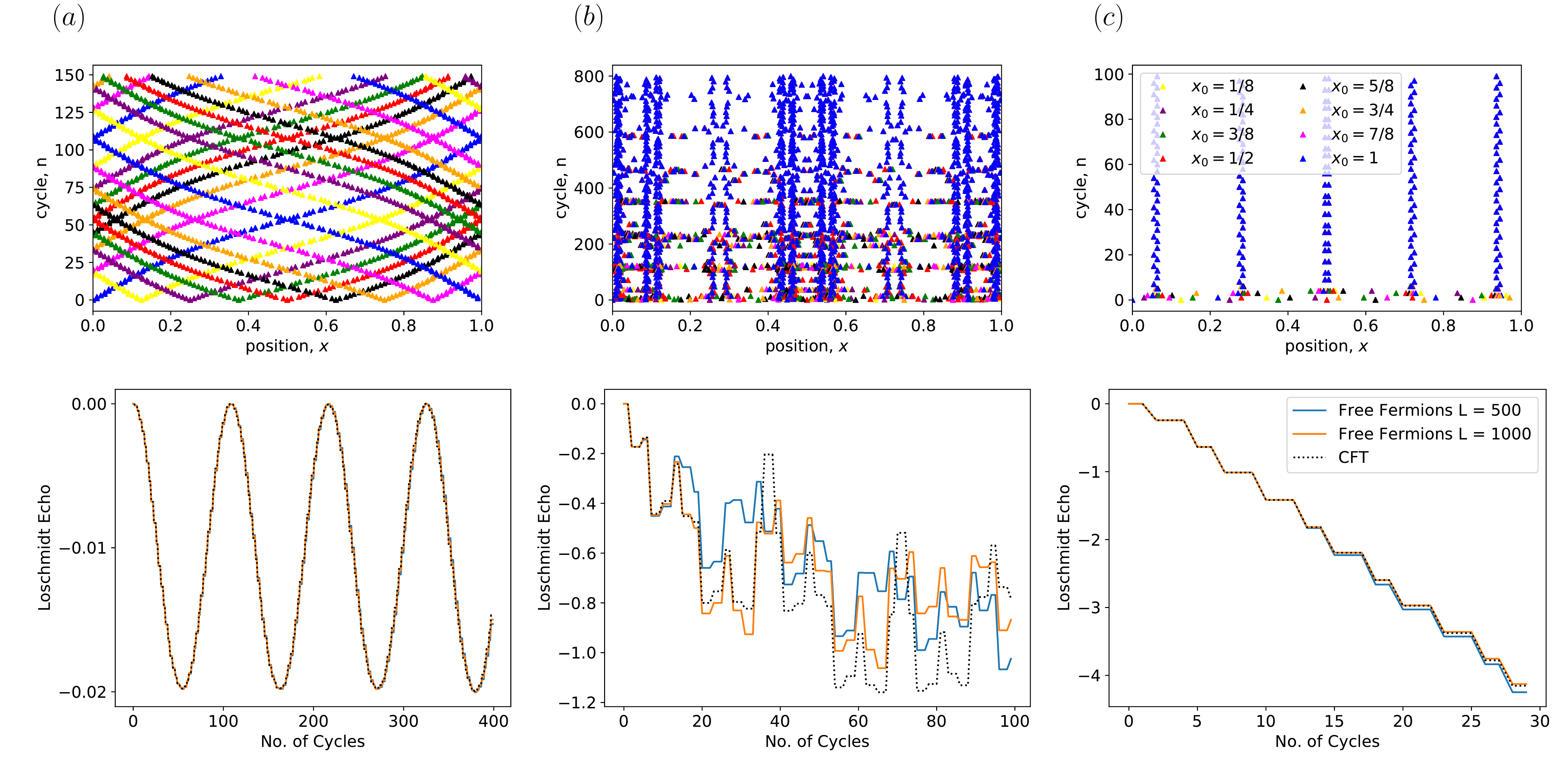}
	\caption{\label{fig:quasiparticlepicture} Top: Stroboscopic quasiparticle evolution as a function of both space and time predicted by the CFT for a variety of initial conditions. $(a)$ High frequency phase, $T/L=0.01$, where the quasiparticle excitations evolve ballistically in an effective curved space-time $(b)$ Fractal region, $T/L=0.55101$.   Beyond a transient time of approximately 400 stroboscopic steps, the quasiparticle trajectories collapse onto a a unique trajectory  independent of  the initial conditions. The quasiparticles stroboscopically explore more and more regions of space as one approaches the Cantor line. $(c)$ High Lyapunov region, $T/L=0.78$.  Here, for all initial conditions and after very few time steps  the time evolution collapses onto a single trajectory alternating between a few fixed points.  Bottom:  A comparison between CFT prediction for the Loschmidt echo  (Eq.~\eqref{loschechoeq}) for the quasiperiodically driven CFT and the analogous quantity for free fermions hopping on a lattice, given by Eq.~\eqref{lochnum}, for the three different regimes. In the high-frequency regime $(a)$ both completely agree. In the strongly heating $(c)$ regime they agree for a large number of stroboscopic steps. In the slow-heating regime $(b)$, because of the fractal structure of the phase diagram the numerical results strongly depend on the number of lattice sites.}
\end{figure*}

We find three representative behaviours 
 depending on the values of $T=\widetilde{T}$. In the high frequency regime where the flow is periodic and does not have fixed points and  the Lyapunov exponent $\mathcal{L} \sim 0$, the excitations evolve periodically  in time akin to the energy for accessible time scales  as seen on Fig.~\ref{fig:quasiparticlepicture}(a).  In the opposing regime of large  $\mathcal{L}$,  we see that after a few  cycles of the quasiperiodic drive, the trajectories of the excitations collapse onto  a unique trajectory independent of the choice of  initial conditions, leading to a localization effect. This trajectory   alternates between a finite number of fixed points at stroboscopic times, as seen on Fig.~\ref{fig:quasiparticlepicture}(c). Such fixed points of the stroboscopic trajectories of the quasiparticles are given by the flow of $\tilde{z}_n$ on Fig.~\ref{fig:energypluslosch}(c).  This is similar to the coherent heating phase of the periodic drive, for which the excitations at stroboscopic times  localize at two points in space, understood as horizons. The energy grows exponentially  with time at these fixed points, whose positions depend on the parameters of the drive.

However, in the fractal region of the phase diagram for the quasiperiodic drive, Fig.~\ref{fig:quasiparticlepicture}(b), the situation is more complex.
As before,  the propagation of the excitations  depends on the initial conditions  upto a transient time $t_0$, and the total energy in the system  does not increase significantly. After this transient  period, the excitations all follow the same trajectory independently of their initial position and localize, but the principal difference with the high Lyapunov case is the manifestation of a  large number of recurring points at stroboscopic times see Fig.~\ref{fig:quasiparticlepicture}(b).  As one approaches  a  point in the Cantor set where $ \mathcal{L}= 0$ the flow diagram  Fig.~\ref{fig:quasiparticlepicture}(b)  is densely filled. Here,  we expect   this transient time $t_0 \to \infty$ concomitant with very slow increase of the energy.  This slow dynamical evolution can be understood by noting that at the Cantor points the trace map remains bounded  for all times.   Since $\text{Tr} (M_n)^2$ also remains bounded, the energy  given by \eqref{lyapuvsenergy}, cannot diverge at Fibonacci steps. To summarise,  we see that by tuning the parameters of the drive $T$ and $\widetilde{T}$ we can encounter regions with fast heating, with a localization of the excitations, as well as regions with very slow heating, where excitations remain delocalized for large times, as opposed to a purely random drive.

\section{Numerical results}
\label{Sec:numerics}

In this section we present numerical results on the free fermion chain for the Loschmidt echo. The quasiperiodic drive is induced by the two Hamiltonians given by
\begin{equation}
\begin{cases}
H=\frac{1}{2}\sum_{i=1}^{L-1}c_i^{\dagger}c_{i+1}+h.c.,\\
\widetilde{H}=\sum_{i=1}^{L-1}\sin^2\left(\frac{\pi(i+\frac{1}{2})}{L}\right)c_i^{\dagger}c_{i+1}+h.c.,
\end{cases}
\label{lochnum}
\end{equation}
where $c_i$ and $c_i^{\dagger}$ are fermionic operators satisfying the usual anticommutation rules. 
Then, following the strategy of Ref.~\onlinecite{wen2018floquet}, one can get the stroboscopic time evolution under the quasiperiodic drive, starting from the ground state with open boundary conditions $|G\rangle$. The Loschmidt echo can then be computed numerically, $\text{F}(NT)=|\langle G|U(NT)|G\rangle|^2$, and we compare it to the CFT prediction, given by Eq.~\eqref{loschechoeq}.


The explicit comparison is shown in Fig.~\ref{fig:quasiparticlepicture}. In the high-frequency domain, Fig.~\ref{fig:quasiparticlepicture}(a) as well as in the high Lyapunov region, Fig.~\ref{fig:quasiparticlepicture}(c), the agreement between the CFT predictions and the free fermion numerics is remarkable for a large number of steps. In the fractal region, characterized by low Lyapunov exponent, the agreement is less striking as observed on Fig.~\ref{fig:quasiparticlepicture}(b). Indeed in this region the Loschmidt echo scales very differently depending on the size of the system, making the explicit comparison with the CFT more complicated, even though the overall scaling of the Loschmidt echo is correctly captured by the CFT. This strong dependence on the system size in the fractal region of the phase diagram can be explained by the strong dependence of the Lyapunov exponent with $T$. Changing the size of the system $L$ while keeping $T/L$ fixed has the effect of redefining $T$. On the lattice such a redefinition can lead to non negligible changes in the Lyapunov exponent in the fractal region, changing the scaling of the Loschmidt echo.\\
The CFT and the free fermion chain are in good agreement for the total energy $E(t=NT)$ in the high-frequency regime, as the energy only oscillates and does not grow exponentially. However as long as the system starts to heat up, the description of the CFT might deviate at long times as it describes the low energy sector of the free fermion chain.

\section{Discussion and conclusions}
We studied the dynamics of quasiperiodically driven CFTs wherein the unitary evolution operator consists of undeformed and {\it{sine-square-deformed}} unitaries repeating quasiperiodically according to the Fibonacci recurrence relation. While on the one hand, it is known  that the periodically driven counterpart of the current setup has a rich phase diagram exhibiting both heating and non-heating phases, the completely random driving generically heats the system up. Therefore our work embodies a natural middle ground between these periodically and randomly driven scenarios. Naively, it seems that such quasiperiodically driven CFTs also generically heat up except for isolated lines in the parameter space of the model as can be seen from the infinite time expectation value of the energy density as well as the positivity of the Lyapunov exponent. We find that the infinite time observables miss out rather rich dynamical phenomena which can be used to distinguish different regions in the parameter space. More precisely, by tuning $T/L$ and $\widetilde{T}/L$, one can go between very slow and fast heating. We distinguish between three distinct scenarios:
\begin{itemize}
\item The non-heating high frequency regime  where the system displays periodic
dynamics with no heating for physically relevant timescales. 
\item The fast heating regimes with large Lyapunov
exponents where one sees the  indefinite  build up of energy at a finite number of points. These points correspond to a small number of fixed points under the flow generated by conformal transformation obtained from the quasiperiodic unitary.
\item The so-called fractal regime, which exists in the neighborhood the non-heating lines (along which the Lyapunov exponent vanishes) in the parameter space. In the neighborhood of these non-heating lines, the dynamics are much slower as compared with the fast-heating regime and in fact the system remains non-heating for experimentally accessible as well as physically relevant timescales.
\end{itemize}    
Our analysis leverages the analytic power of CFT on the one hand and certain rich mathematical structures which have been historically used to analyze quasi-crystals on the other. Therefore such a model is one of the few solvable examples of a driven quantum many body system where one can tune between regimes with different heating rates. Another feature of our setup is that it doesn't inherently rely on interactions or on disorder. 

\medskip \noindent It is worth mentioning that the fractal structure is not contingent on the particular choice of deformation and we expect such features to survive more generic deformations\cite{moosavi2019inhomogeneous}. In Appendix~\ref{Sec:appendixD} we propose a first generalization of those results to generic M\"obius deformations of the Hamiltonian density. However the choice of sine-square deformation and its inherent $SL(2,\mathbb C)$ structure makes the connection with $1$-dimensional quasi-crystals more natural.

\medskip \noindent There are several rich and interesting directions to pursue. These may involve generalizing the sine-square deformation to other possible kinds of deformations, studying other diagnostics of heating and thermalization within our setup, understanding operator scrambling and chaos in driven CFTs to name a few.

{\it Note added} --- During the preparation of this manuscript, we learnt about Ref.~{\onlinecite{quasiperiodicwen}}
which appears in the same arXiv posting and also discusses similar
results regarding Fibonacci quasi-periodically driven
CFTs. We
thank the authors for sending us their manuscript before
posting it online.
\section*{Acknowledgements}
We would like to thank A. Rosch for discussions.
This project has received funding from the European Research Council (ERC) under the European Union's Horizon 2020 research and innovation program (ERC-StG-Neupert-757867-PARATOP and Marie Sklodowska-Curie grant agreement No 701647).

\bibliography{quasiperiodic_bib}

\clearpage
\begin{widetext}

\appendix
\section{Loschmidt echo}
\label{applosch}
In this section we derive the expression for the Loschmidt echo $\text{F}(t)$ during the quasiperiodic drive,
\begin{equation}
\text{F}(t)=|\langle\psi(0)|\psi(t)\rangle|^2.
\end{equation}
Let us first compute the Loschmidt echo in the setup considered in Ref.~\onlinecite{Wen_2018} of a single quench with the SSD Hamiltonian at $t=0$, starting from a generic excited state $|\Phi\rangle$ of $\mathcal{H}$. The Loschmidt echo is then
\begin{equation}
\text{F}(t)=|\langle \Phi|\ee^{-\ii \widetilde{\mathcal{H}}t}|\Phi\rangle|^2
\end{equation}
The state $|\Phi\rangle$ can always be written as an in-state generated by a primary field $\phi(z,\bar{z})$ of conformal weights $(\Delta,\bar{\Delta})$ acting on the $SL(2,\mathbb{C})$ invariant vacuum $|0\rangle$ at $\tau\rightarrow-\infty$, which corresponds to inserting the field at the origin of the complex plane after applying the exponential mapping in the $z$ coordinates,
\begin{equation}
|\Phi\rangle=\lim_{z,\bar{z}\rightarrow 0}\phi(z,\bar{z})|0\rangle.
\end{equation}
The computation then reduces to
\begin{equation*}
\langle \Phi|e^{-\widetilde{H}\tau}|\Phi\rangle = \lim_{z_1, \bar{z}_1\rightarrow 0}\lim_{z_2, \bar{z}_2\rightarrow 0} z_2^{-2\Delta}\bar{z}_2^{-2\bar{\Delta}} \langle 0|\phi\left(z_2^{-1},\bar{z}_2^{-1}\right) e^{-\widetilde{H}\tau}\phi(z_1,\bar{z}_1) |0\rangle.
\end{equation*}
We now insert the identity $\mathbb{I}=\ee^{\widetilde{\mathcal{H}}\tau}\ee^{- \widetilde{\mathcal{H}}\tau}$, and use the fact that $|0\rangle$ is an eigenstate of $\widetilde{\mathcal{H}}$, as $L_0|0\rangle=L_{\pm1}|0\rangle=0$, therefore $\ee^{- \widetilde{\mathcal{H}}\tau}$ acting on $|0\rangle$ gives a phase irrelevant for the Loschmidt echo. By going to the $\tilde{z}_0(z)$ coordinates, explicitly given by \eqref{tildezeq}, we obtain
\begin{equation}
\langle \Phi|e^{-\widetilde{H}\tau}|\Phi\rangle = \lim_{z_1, \bar{z}_1\rightarrow 0}\lim_{z_2, \bar{z}_2\rightarrow 0}\left(\frac{\partial \tilde{z}_0}{\partial z}\bigg|_{z_1}\right)^{\Delta}\left(\frac{\partial \bar{\tilde{z}}_0}{\partial \bar{z}}\bigg|_{\bar{z}_1}\right)^{\bar{\Delta}} z_2^{-2\Delta}\bar{z}_2^{-2\bar{\Delta}} \langle 0|\phi\left(z_2^{-1},\bar{z}_2^{-1}\right) \phi(\tilde{z}_0(z_1),\tilde{\bar{z}}_0(\bar{z}_1)) |0\rangle.
\end{equation}
Finally, $\langle 0|\phi\left(z_2^{-1},\bar{z}_2^{-1}\right) \phi(\tilde{z}_0(z_1),\tilde{\bar{z}}_0(\bar{z}_1)) |0\rangle$ is a simple two point function in the $\tilde{z}$ coordinates, leading to
\begin{equation}
\langle \Phi|e^{-\widetilde{H}\tau}|\Phi\rangle=\lim_{z_1,\bar{z}_1\rightarrow 0}\lim_{z_2,\bar{z}_2\rightarrow 0} \left(\frac{\partial \tilde{z}_{0}}{\partial z}\bigg|_{z_1}\right)^{\Delta}\left(\frac{\partial \bar{\tilde{z}}_{0}}{\partial \bar{z}}\bigg|_{\bar{z}_1}\right)^{\bar{\Delta}}\frac{z_2^{-2\Delta}}{\left|z_2^{-1}-\tilde{z}_{0}(z_1)\right|^{2\Delta}}\frac{\bar{z}_2^{-2\bar{\Delta}}}{\left|\bar{z}_2^{-1}-\bar{\tilde{z}}_{0}(\bar{z}_1)\right|^{2\bar{\Delta}}}.
\end{equation}
The limits can then be taken, giving the same result independently of their order,
\begin{equation}
\lim_{z_1,z_2\rightarrow 0}\frac{z_2^{-2\Delta}}{\left|z_2^{-1}-\tilde{z}_{0}(z_1)\right|^{2\Delta}}=\lim_{z_2\rightarrow 0}\frac{z_2^{-2\Delta}}{|z_2^{-1}-\alpha|^{2\Delta}}=1.
\end{equation}
Therefore only the derivative terms contribute, whose limit give
\begin{equation}
\lim_{z_1\rightarrow 0}\frac{\partial \tilde{z}_{0}}{\partial z}\bigg|_{z_1}=\frac{1}{(1-\frac{\pi \tau}{L})^2}.
\end{equation}
The same steps apply for the anti-holomorphic part. Therefore the final result for a single quench with the SSD Hamiltonian $\widetilde{\mathcal{H}}$ after analytic continuation to real times $\tau\rightarrow \ii t$ is 
\begin{equation}
\text{F}(t)=\frac{1}{\left(1+\frac{\pi^2t^2}{L^2}\right)^{2(\Delta+\bar{\Delta})}}.
\end{equation}
Leading to an quadratic decay of the Loschmidt echo during the quench. In the case of a periodic drive, one simply needs to replace $\tilde{z}_0(z)$ by $\tilde{z}_n(z)=\tilde{z}_0\circ...\circ \tilde{z}_0(z)$, which can be expressed explicitly in terms of the normal form of the 1-cycle transformation to obtain that the Loschmidt echo decays exponentially in the heating phase, quadratically at the phase transition, and oscillates in time in the non-heating phase, leading to periodic revivals in the system. In case of the quasiperiodic drive, writing the transformation after $n$ steps $\tilde{z}_n(z)=\frac{\alpha_nz+\beta_n}{\gamma_n z+\delta_n}$, we obtain that the Loschmidt echo is simply
\begin{equation}
\text{F}(t)=\left|\frac{\alpha_n\delta_n-\beta_n\gamma_n}{\delta_n^2}\right|^{2(\Delta+\bar{\Delta})}.
\end{equation}

\section{Fibonacci trace map} 
\label{rouge}
The quasiperiodicity induced by a Fibonacci sequence has already been studied in the context of one dimensional quasicrystal literature \cite{PhysRevLett.50.1870,PhysRevB.35.1020,PhysRevB.61.6645}. In this section we focus on the Fibonacci trace map approach to finding the spectrum of such a Hamiltonian. The tight-binding Hamiltonian describing the quasiperiodic Fibonacci chain is\cite{1987CMaPh.111..409S}
\begin{equation}
H_{mn}=\delta_{m,n+1}+\delta_{m+1,n}+\delta_{m,n}\lambda v(n),
\label{specfibo}
\end{equation}
where $v(n)=|\nu(n-1)-1|$.
The one dimensional Fibonacci quasicrystal is described by the Schrodinger equation
\begin{equation}
\psi_{n-1}+\psi_{n+1}+v(n)\psi_n=E\psi_n.
\end{equation}
The Schrodinger equation can be rewritten using transfert matrices as:
\begin{equation}
\begin{bmatrix}
\psi_{n+1} \\ \psi_{n}
\end{bmatrix}
=
T(n)\begin{bmatrix}
\psi_n \\ \psi_{n-1}
\end{bmatrix}
=
\begin{bmatrix}
E-v(n) & -1\\
1 & 0
\end{bmatrix}
\begin{bmatrix}
\psi_n \\ \psi_{n-1}
\end{bmatrix}
\end{equation}
This can be iterated, such that finding the eigenvectors $\psi_n$ amounts to finding the product of $n$ matrices $T(n)$. We now define $T_j=T(F_j)...T(1)$. Then it is straightforward to show that for $T_j\in SL(2,\mathbb{C})$, the following recursion holds: $T_{j+1}=T_{j-1}T_j$. This relation can be rewritten in terms of the trace of the matrices
\begin{equation}
\text{Tr}(T_{j+1})=\text{Tr}(T_j)\text{Tr}(T_{j-1})-\text{Tr}(T_{j-2}).
\end{equation}
Therefore writing $x_j=\frac{1}{2}\text{Tr}(T_j)$, the Fibonacci trace map reduces to
\begin{equation}
x_{j+1}=2x_jx_{j-1}-x_{j-2}.
\end{equation}
We also introduce $y_j=\frac{1}{2}\text{Tr}(T_{j+1})$ and  $z_j=\frac{1}{2}\text{Tr}(T_{j+2})$. One can then define a discrete dynamical map $\mathcal{T}$
\begin{equation}
\mathcal{T}:\mathbb{R}^3\rightarrow \mathbb{R}^3,\quad (x_i,y_i,z_i)\mapsto (y_i,z_i,2y_iz_i-x_i).
\end{equation}
This mapping has been introduced in the quasicrystal literature as the \textit{Fibonacci trace map}. The mathematical structure of such a mapping has been studied in e.g Ref.~{\onlinecite{1987CMaPh.111..409S, Kadanoff1276,yessen2015newhouse,PhysRevLett.50.1870,Damanik_2016,BAAKE_1993}}. One crucial property of the Fibonacci trace map is that it admits an invariant $I(x_j,y_j,z_j)$:
\begin{equation}
I(x_j,y_j,z_j)=x_j^2+y_j^2+z_j^2-2x_jy_jz_j-1,
\end{equation}
which is the same for any $j\in\mathbb{N}$. Then it is natural to consider the cubic level surfaces $S_V=\{(x,y,z)\in\mathbb{R}^3 \mid I(x,y,z)=V\}$. The surface $S_V$ has different topologies depending on the sign of $V$. The manifold is noncompact for $V>0$. At $V=0$ the middle part of the manifold is compact and touches the other 4 non compact components at a single point. For $V<0$ the middle part is completely detached from the 4 noncompact components.
In the case of the Fibonacci quasicrystal, the invariant is always positive and given by $V=\frac{\lambda^2}{4}\geq 0$. One can then study the set of bounded orbits under infinite iterations of the Fibonacci trace map, starting from the initial point $(1,\frac{E}{2},\frac{E-\lambda}{2})$.
In order to make the connection between the orbits of the Fibonacci trace map and the spectrum $E$ of the Fibonacci Hamiltonian, the following theorem has been proved in Ref.~{\onlinecite{1987CMaPh.111..409S}}:
\begin{prop}
An energy $E\in\mathbb{R}$ belongs to the spectrum of the discrete Fibonacci Hamiltonian if and only if the positive semiorbit of the point $(1,\frac{E}{2},\frac{E-\lambda}{2})$ under iterates of the trace map $\mathcal{T}$ is bounded.
\end{prop}
Therefore finding the set of bounded orbit under the Fibonacci trace map is completely sufficient to determine the spectrum of the Fibonacci Hamiltonian. The following theorem enables to conclude that the spectrum of the Fibonacci quasicrystal is a fractal set similar to a Cantor set. It has been first proved for $\lambda\geq4$ in Ref.~{\onlinecite{1987CMaPh.111..409S}}, and then for any $\lambda>0$ in Ref.~{\onlinecite{sutoandras}}:
\begin{prop}
The set of bounded orbits is a Cantor set for $\lambda>0$.
\end{prop}
The spectrum $E$ of \eqref{specfibo} has therefore a fractal structure similar to the Cantor set, and is of measure 0.

\begin{figure}[htb]
	\includegraphics[width=8.6cm]{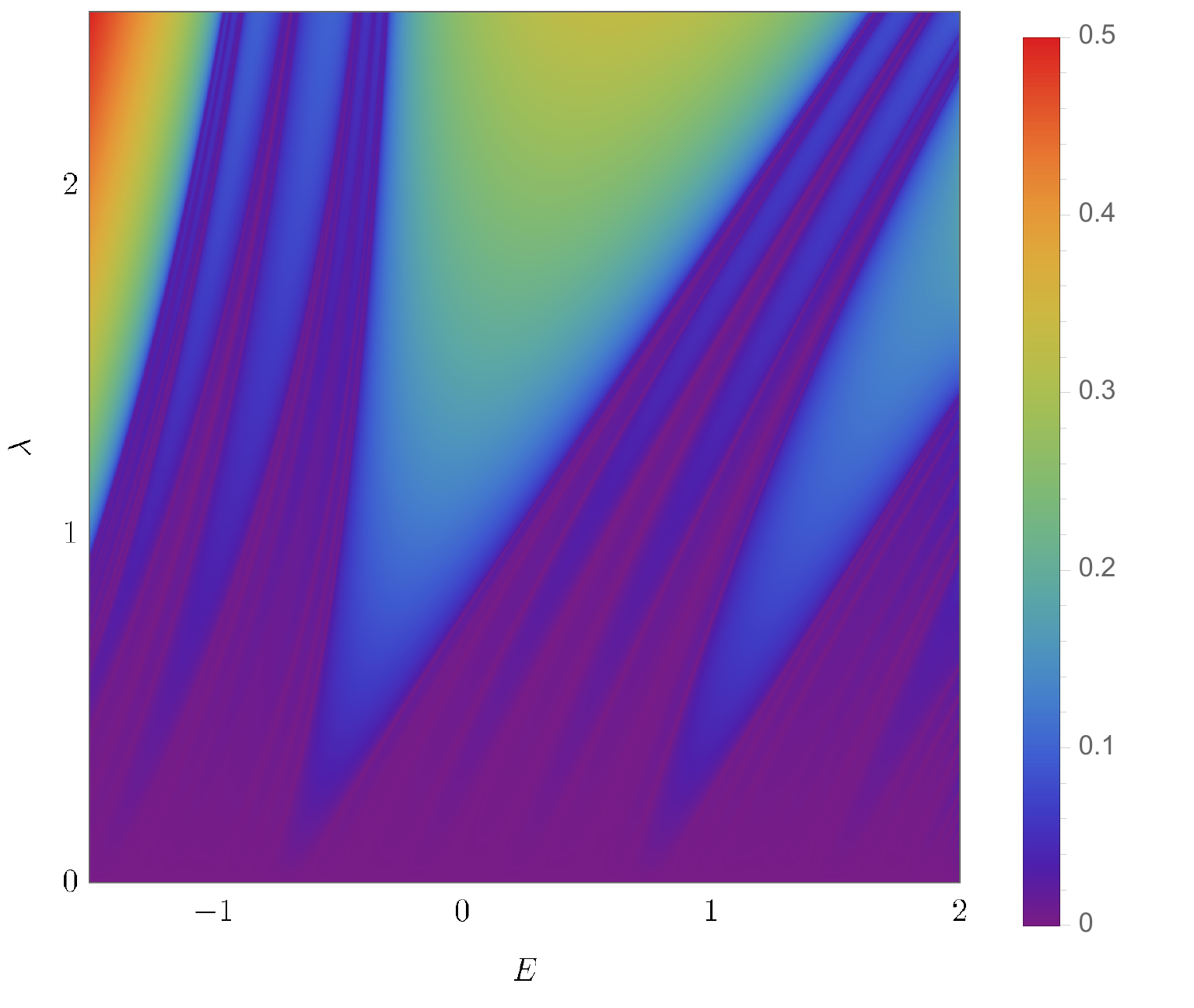}
	\caption{\label{fig:spectrumfibo} Lyapunov exponent as a function of the spectrum $E$ and the parameter $\lambda$, using the correspondence \eqref{correspondence}. This Figure exactly reproduces the Figure for the spectrum $E(\lambda)$ of the Fibonacci quasicrystal, which can be found in Ref.~\onlinecite{damanik2010}. Here, regions of non-zero Lyapunov exponent are dense for $\lambda>0$ and correspond to gaps in the spectrum. The spectrum itself consists in regions of $0$ Lyapunov exponent, of measure zero.}
\end{figure}

\section{High frequency expansion}
\label{highfreq}
In this section we give some details of the derivation of the effective stroboscopic Hamiltonian in the high-frequency approximation, mostly relying on the strategy of Ref.~\onlinecite{PhysRevB.99.020306}.
Consider the quasiperiodic drive defined by Eq.~\eqref{timeevofibo}. In this case, at the step $n$, there are $F_{n+1}$ unitary operators, corresponding to either the SSD evolution for time $\wt{T}$ or uniform evolution for time $T$. Assuming that $T=\wt{T}\ll L$, one can then make the approximation that
\begin{equation}
\ee^{-\ii T\mathcal{H}}\ee^{-\ii T\wt{\mathcal{H}}}=\ee^{-\ii T(\mathcal{H}+\wt{\mathcal{H}}+ \ii\frac{T}{2}[\mathcal{H},\wt{\mathcal{H}}])}
\end{equation}
Then, as one considers more unitary operators, we will still only keep the first order commutators. Therefore at step $n$, there will be $F_{n+1}$ unitary, among which there will be $F_{n}$ times $\ee^{-\ii \mathcal{H}T}$ and $F_{n-1}$ times $\ee^{-\ii \wt{\mathcal{H}}T}$, using that $F_{n+1}=F_n+F_{n-1}$. Generalizing this strategy to any stroboscopic step $N$ and not only to $N=F_{n+1}$, one needs to introduce the binary function $\nu(N)\in\{0,1\}$ defined in the main text. If $\nu(N)\in \{0,1\}$ is $1$, the unitary $\ee^{-\ii \mathcal{H}T}$ appears at step $N$, and if $1-\nu(N)\in\{0,1\}$ is $1$, the unitary $\ee^{-\ii \wt{\mathcal{H}}T}$ appears at step $N$. Therefore the numbers of unitary appearing up to the step $N$ is given by $\rho(N)$ and $\sigma(N)$ defined in Eq.~\eqref{somedefini}. The final step consists in counting how many commutators $[\mathcal{H},\wt{\mathcal{H}}]$ and $[\wt{\mathcal{H}},\mathcal{H}]$ appear, given by $\tau(N)$ in Eq.~\eqref{somedefini}. 
This leads to the following form of effective Hamiltonian, defined as $U_N=^{-\ii NT \mathcal{H}_{\text{eff}}^{(N)}}$,
\begin{equation}
\mathcal{H}_{\text{eff}}^{(N)}=\frac{\rho(N)}{N}\mathcal{H}+\frac{\sigma(N)}{N}\wt{\mathcal{H}}+\ii\frac{T}{2} \frac{\tau(N)}{N}[\mathcal{H},\wt{\mathcal{H}}].
\end{equation}
Using the fact that the commutator is $[\mathcal{H},\widetilde{\mathcal{H}}]=-\left(\frac{2\pi}{L}\right)^2\frac{1}{2}(L_1-L_{-1})$, the stroboscopic Hamiltonian can be rewritten as
\begin{equation}
\mathcal{H}_{\text{eff}}^{(N)}=\frac{2\pi}{L}\left[\left(\frac{\rho(N)}{N}+\frac{\sigma(N)}{N}\right)L_0+\left(-\frac{\sigma(N)}{2N}-\ii \frac{\pi}{2}\frac{T}{L}\frac{\tau(N)}{N}\right)L_1+\left(-\frac{\sigma(N)}{2N}+\ii \frac{\pi}{2}\frac{T}{L}\frac{\tau(N)}{N}\right)L_{-1}\right].
\end{equation}
Note that this Hamiltonian is still a linear combination of the generators of $SL(2,\mathbb{R})$. It can then be written in the form $\mathcal{H}_{\text{eff}}=\int_0^L\dd x f_N(x) T_{00}(x)$, with the effective deformation given by
\begin{equation}
f_N(x)=\frac{\rho(N)}{N}+\frac{\sigma(N)}{N}-\frac{\sigma(N)}{N}\cos\left(\frac{2\pi x}{L}\right)+\frac{\pi T}{L}\frac{\tau(N)}{N}\sin\left(\frac{2\pi x}{L}\right).
\label{effdef}
\end{equation}
\begin{figure}[htb]
	\includegraphics[width=8.6cm]{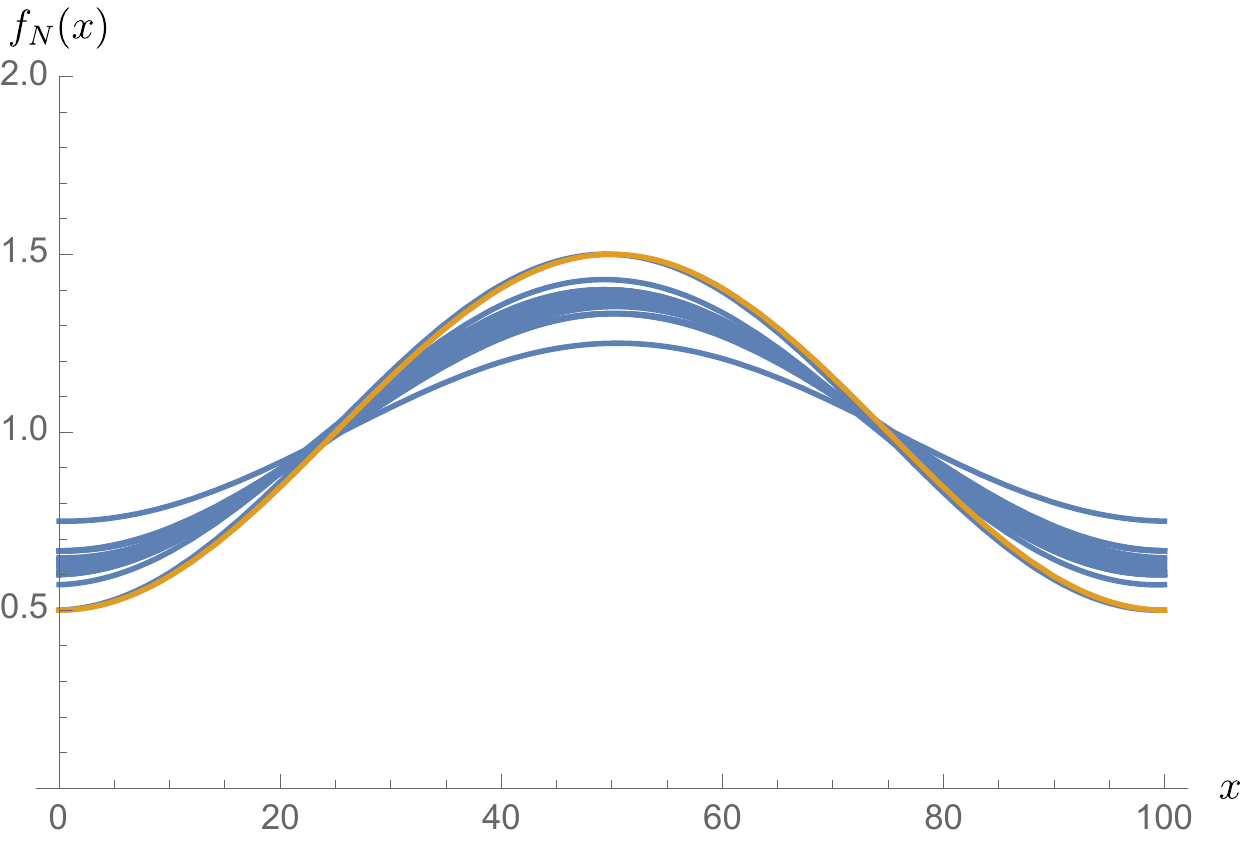}
	\caption{\label{fig:effectivedeformation} Effective deformation $f_N(x)$ given by \eqref{effdef}, for $L=100$, $\frac{T}{L}=0.01$, $N\in\{1,20\}$. The orange curve corresponds to the effective deformation of the periodic drive, $f(x)=1-\frac{1}{2}\cos\left(\frac{2\pi x}{L}\right)$. }
\label{highfreqprof}
\end{figure}

We are also interested in the late time behaviour of the coefficients $\rho$ $\sigma$ and $\tau$. At the step $N=F_{n+1}$, the coefficients are found to scale as
\begin{equation}
\begin{cases}
\rho(N)\sim \frac{\Phi^n}{\sqrt{5}}\\
\sigma(N)\sim \frac{\Phi^{n-1}}{\sqrt{5}}\\
\tau(N)\sim \epsilon\frac{\Phi^{n+1}}{\sqrt{5}}
\end{cases}
\end{equation}
where $\epsilon\in[-\frac{1}{\Phi},1-\frac{1}{\Phi}]$. Therefore one can find that the quadratic Casimir invariant of $SL(2,\mathbb{R})$ will scale as
\begin{equation}
c^{(2)}(N)=\frac{1}{N^2}\left[(\rho(N)+\sigma(N))^2-\sigma(N)^2-\left(\frac{\pi T}{L}\right)^2\tau(N)^2\right]\sim \frac{\Phi^{2n}}{5N^2}\left(1+2\Phi^{-1}-\left(\frac{\pi T}{L}\right)^2\epsilon^2\Phi^2\right)
\end{equation}
To check whether or not this invariant could take negative values, we look at the lower bound $|\epsilon_{\text{min}}|=\frac{1}{\Phi}$, to conclude that the quadratic Casimir is negative at long times if
\begin{equation}
\frac{T}{L}\geq \frac{\sqrt{1+2\Phi^{-1}}}{\pi}\approx 0.47
\end{equation}
Therefore the Casimir invariant at long times has to be positive in the approximation $T\ll L$, meaning that within this approximation heating should not occur at long times. From the Fibonacci trace map point of view, heating would actually occur when the orbit escapes, but that will happen after times which are not physically relevant, and not captured by this first order expansion.

\section{M\"obius quasiperiodic drive}
\label{Sec:appendixD}

In this section we propose to study a whole family of Fibonacci quasiperiodic drives alternating between the uniform Hamiltonian $\mathcal{H}$, and the so-called M\"obius Hamiltonian $\mathcal{H}_{\text{M\"ob}}(\theta)$, introduced initially as a regularisation of the SSD Hamiltonian, and defined as
\begin{equation}
    \mathcal{H}_{\text{M\"ob}}(\theta)=L_0-\frac{\tanh(2\theta)}{2}(L_1+L_{-1})+\overline{L}_0-\frac{\tanh(2\theta)}{2}(\overline{L}_1+\overline{L}_{-1}).
\label{mobiushamiltonian}
\end{equation}
This Hamiltonian interpolates between the uniform Hamiltonian, $\mathcal{H}= \mathcal{H}_{\text{M\"ob}}(0)$ and the SSD Hamiltonian, $\widetilde{\mathcal{H}}= \lim_{\theta\rightarrow \infty}\mathcal{H}_{\text{M\"ob}}(\theta)$. It can be seen as a deformed Hamiltonian of the form of Eq.~\eqref{genericdef}, with $f(x)=1-\tanh(2\theta)\cos\left(\frac{2\pi x}{L}\right)$. Just as for the SSD Hamiltonian, the time evolution with the M\"obius is encoded in a conformal transformation which is a M\"obius transformation. Such a transformation is explicitly given by\cite{Wen_2018}
\begin{equation}
\tilde{z}_{\theta}(z)=\frac{\left[(1-\lambda)\cosh(2\theta)-(\lambda+1)\right]z+(\lambda-1)\sinh(2\theta)}{(1-\lambda)\sinh(2\theta)z+\left[(\lambda-1)\cosh(2\theta)-(\lambda+1)\right]},
\label{eqmobmob}
\end{equation}
where $\lambda=\ee^{\frac{2\pi \tilde{\tau}}{L\cosh(2\theta)}}$. In the limit $\theta\rightarrow\infty$, this M\"obius transformation reduces to Eq.~\eqref{tildezeq}. In particular, it can be normalized by multiplying the associated matrix by a factor or $\frac{1}{2}\lambda^{-1/2}$.\\

\medskip \noindent In the case of finite $\theta$, the dynamics with such a Hamiltonian is periodic with period $L\cosh(2\theta)$, and the associated quadratic invariant $c^{(2)}(\theta)=1-\tanh(2\theta)^2$ is strictly positive, implying that the spectrum of such Hamiltonian is discrete and scales as $(L\cosh(2\theta))^{-1}$. In the limit $\theta\rightarrow \infty$ the period tends to infinity, the invariant goes to $0$ and the spectrum is continuous, corresponding to the SSD limit.\\
We can then study the dynamics of the Fibonacci quasiperiodic drive alternating quasiperiodically between $\mathcal{H}$ and $\mathcal{H}_{\text{M\"ob}}(\theta)$ using the same strategy as presented in Sec.~\ref{Sec:method}, replacing Eq.~\eqref{tildezeq} by Eq.~\eqref{eqmobmob}, and understand if the features present in the case of the SSD quasiperiodic drive still survive in this case.
\begin{figure}[h!]
	\includegraphics[width=18cm]{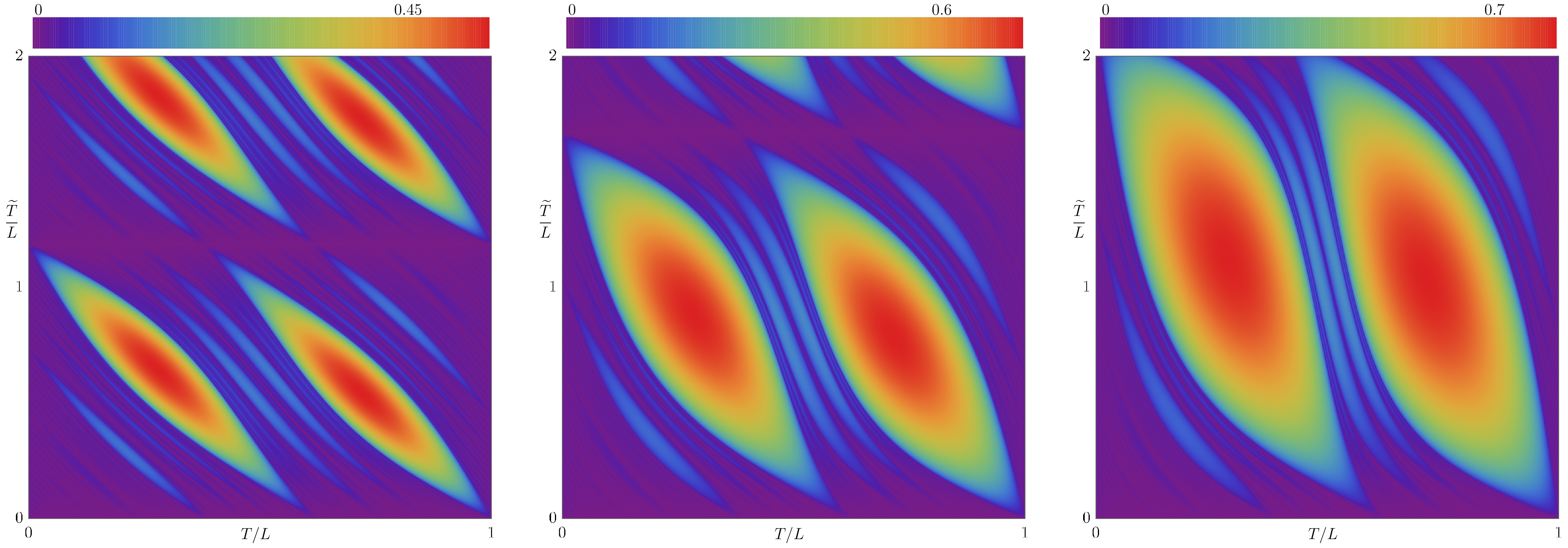}
	\caption{\label{fig:mobdiagram}  M\"obius phase diagrams obtained from the Lyapunov exponent given by Eq.~\eqref{eq:lyapms} for the Fibonacci quasiperiodic drive between $\mathcal{H}$ and $ \mathcal{H}_{\text{M\"ob}}(\theta)$ defined by Eq.~\eqref{mobiushamiltonian} for $\theta=0.3$, $\theta=0.55$, $\theta=0.7$.  }
\label{mobdiagram}
\end{figure}
The resulting phase diagrams for a few choices of $\theta$ are shown in Fig.~\ref{mobdiagram}.	We observe that the phase diagrams are periodic both in $T/L$ as well as $\widetilde{T}/L$ directions, in contrast with the SSD quasiperiodic drive which is only periodic in the $T/L$ directions, as the periodicity induced by the SSD Hamiltonian is infinite, which is recovered by taking the limit $\theta\rightarrow\infty$. These phase diagrams also display an emergent fractal structure where the Lyapunov exponent takes arbitrary small values. The M\"obius deformations remain within the $SL(2,\mathbb{C})$ subalgebra of the Virasoro algebra for arbitrary $\theta$, and further study is needed to determine whether such a fractal structure can be obtained for any general spatial deformation $f(x)$ of the Hamiltonian density, for which we cannot rely on the Fibonacci trace map, inherent to the $SL(2,\mathbb{C})$ structure of the problem. We also note that one can compute the invariant of the Fibonacci trace map associated, which is
\begin{equation}
    I=\sin^2\left(\frac{\pi T}{L}\right)\sin^2\left(\frac{\pi \widetilde{T}}{L}\text{sech}(2\theta)\right)\sinh(2\theta)^2.
\end{equation}
It is straightforward to verifiy that in the SSD limit one recovers the invariant \eqref{invarinv}. Once again the invariant is positive for any choice of driving parameters, and therefore the set of bounded orbits under the Fibonacci trace map still forms a fractal set, which will get denser as $\theta\rightarrow 0$. We note that the explicit mapping to the Fibonacci quasi-crystal, given by Eq.~\eqref{correspondence} for the SSD quasiperiodic drive, cannot be explicitly found in the case of finite $\theta$.

\end{widetext}
\end{document}